\documentclass[useAMS,usenatbib]{mn2e}

\pdfoutput=1
\usepackage{vector}  
\usepackage{bm}
\usepackage{latexsym,amssymb,amsmath} 
\usepackage{amsmath, amsthm, amssymb}
\usepackage{graphicx}

\title[Bayesian Computation Methods - A Comparison]{Bayesian Computational Methods - A Comparison}
\author[A. Tua and K. Zarb Adami]{A. Tua$^{1}$\thanks{E-mail: alan.tua@cern.ch} and K. Zarb Adami$^{2}$\thanks{E-mail: kza@astro.ox.ac.uk}\\
$^{1}$University of Malta, Malta\\
$^{2}$University of Oxford, England}

\begin{document}

\date{Submitted to arxiv.org in March 2010}

\pagerange{\pageref{firstpage}--\pageref{lastpage}} \pubyear{2010}

\maketitle

\label{firstpage}

\begin{abstract}
This paper focuses on utilizing two different Bayesian methods to deal with a variety of toy problems which occur in data analysis. In particular we implement the Variational Bayesian and Nested Sampling methods to tackle the problems of polynomial selection and Gaussian Mixture Models, comparing the algorithms in terms of processing speed and accuracy. In the problems tackled here it is the Variational Bayesian algorithms which are the faster though both results give similar results.
\end{abstract}

\begin{keywords}
Bayesian, Variational Bayes, Nested Sampling, Evidence
\end{keywords}

\section{Introduction}
Parameter estimation within the Bayesian framework rests on the application of Bayes theorem to data analysis. If we consider a parameter set $\theta$, a data set $\mathbb{D}$ and all prior knowledge $\mathbb{I}$ Bayes theorem tells us that:
\begin{equation}
\label{BAYES}
P(\theta|\mathbb{D},\mathbb{I}) = \frac{L(\mathbb{D}|\theta,\mathbb{I})P(\theta|\mathbb{I})}{P(\mathbb{D}|\mathbb{I})}
\end{equation}
Here we have:
\begin{itemize}
\item The \emph{Posterior Probability} $P(\theta|\mathbb{D},\mathbb{I})$ which represents our belief in the hypotheses once we have analysed the data available.
\item The \emph{Prior Probability} $P(\theta|\mathbb{I})$ which encodes our previous knowledge of the system under examination. 
\item The \emph{Likelihood Function} $L(\mathbb{D}|\theta,\mathbb{I})$ which is the probability of observing the data for a given set of parameters.
\item The \emph{Evidence} $P(\mathbb{D}|\mathbb{I})$ which is a normalisation constant given by the probability of the data. 
\end{itemize}
In the specific case when the problem being faced is one of parameter estimation we can ignore the denominator which is independent of $\theta$. We thus get:
\begin{equation}
\label{BAYESPROP}
P(\theta|\mathbb{D},\mathbb{I}) \propto L(\mathbb{D}|\theta,\mathbb{I})P(\theta|\mathbb{I})
\end{equation}
Here the prior knowledge is being transformed by the data through the likelihood to give the posterior distribution which embodies our new beliefs. 

Bayesian methodology can also be applied to model selection scenarios to choose between competing hypotheses or models $M_i$. Given a model $M$ we can write down, using Bayes theorem, the probability that $M$ is the correct model given the data $\mathbb{D}$ and our previous knowledge $\mathbb{I}$:
\begin{equation}
P(M|\mathbb{D},\mathbb{I}) = \frac{P(\mathbb{D}|M,\mathbb{I})P(M|\mathbb{I})}{P(\mathbb{D}|\mathbb{I})}
\end{equation}
This is analogous to Equation (\ref{BAYESPROP}) which deals with parameter estimation: just as the posterior probability distribution function for a parameter is proportional to its prior times its likelihood, so the posterior probability for a model as a whole is proportional to its prior probability times its Evidence. $P(\mathbb{D}|\mathbb{I})$ itself cannot be calculated but model selection can still be performed by comparing the probability of two models $M_1$ and $M_2$ using a ratio, called the \emph{posterior ratio} or \emph{odd's ratio}:
\begin{eqnarray}
\label{MODEL_SELECTION}
\frac{P(M_1|D,I)}{P(M_2|D,I)} = \frac{P(M_1|I)}{P(M_2|I)}\times\frac{P(D|M_1,I)}{P(D|,M_2,I)}
\end{eqnarray}
When we have no prior preference for one model over another Equation (\ref{MODEL_SELECTION}) reduces to a ratio of Evidences. Marginilisation and Bayes theorem allow us to express the Evidence as an integral over the parameter space:
\begin{equation}
\label{equ:Evidence}
P(\mathbb{D}|M,\mathbb{I})=\int_{\theta\in\Omega}{P(\theta|M,\mathbb{I})L(\mathbb{D}|\theta,M,\mathbb{I}) d\theta}
\end{equation}
Calculating this integral then allows the experimentalist to select the optimum model to describe a data set. We now discuss two different techniques which lead to a value the Evidence and further on compare them in terms of computational speed and accuracy.

\section{Nested Sampling}
\begin{figure}
  \centering
    \includegraphics[width=3in]{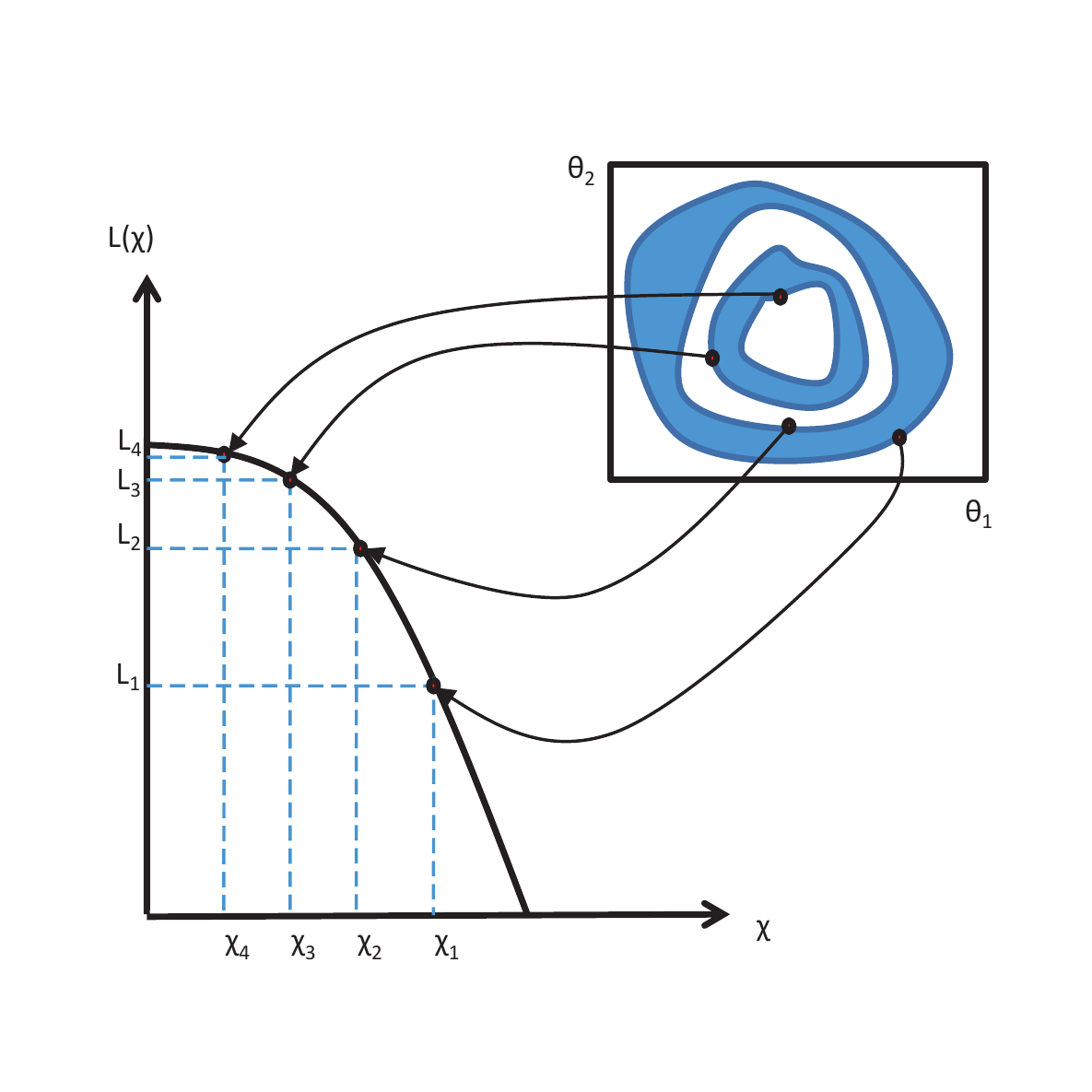}
  \caption{The above diagram, taken from (Skilling, 2006) shows how points in the parameter space are sampled such that they satisfy the Likelihood constraint $L>L^*$.}
  \label{fig:contours}
\end{figure}

\begin{figure}
  \centering
    \includegraphics[width=2.3in]{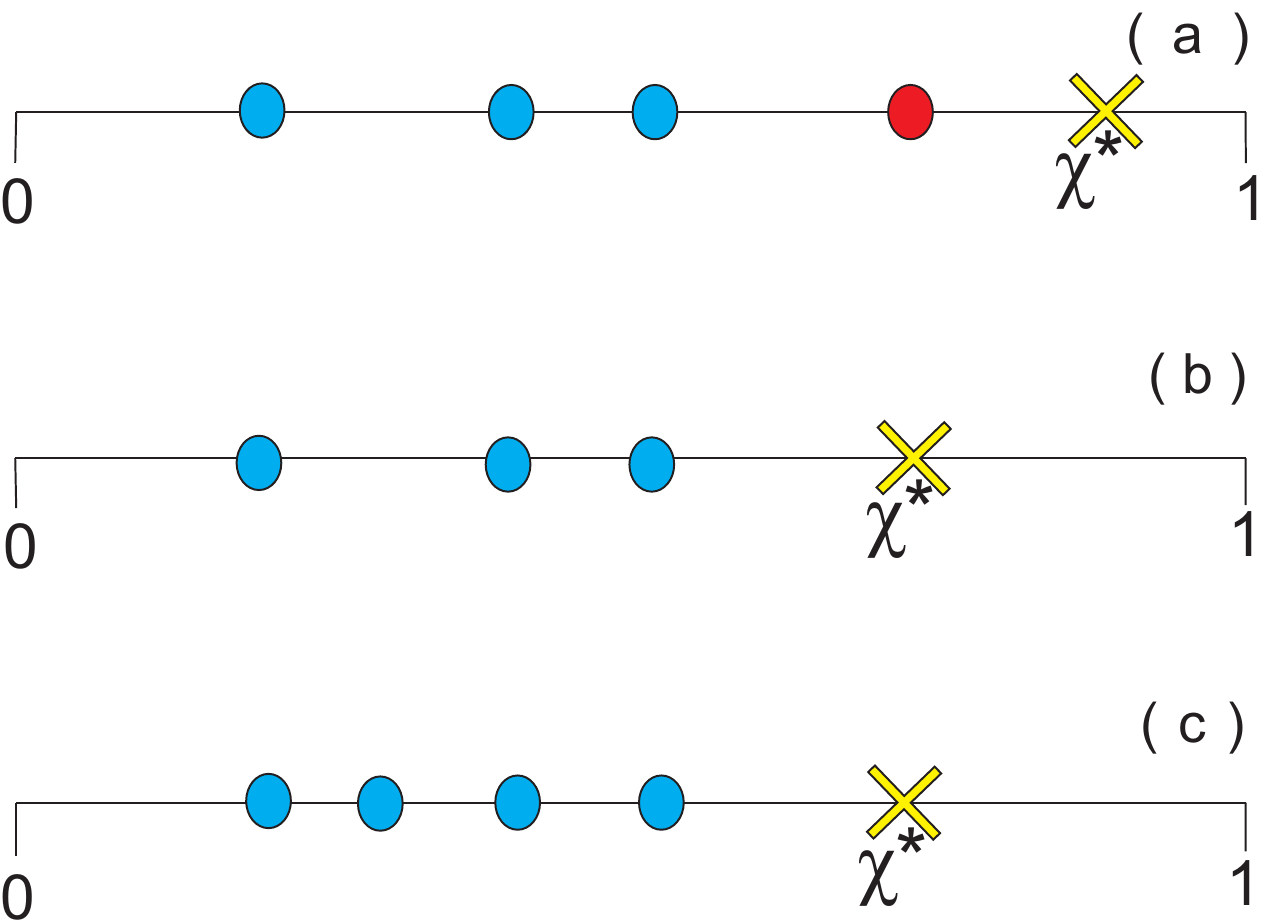}
  \caption[Nested Sampling Iterations]{We see the separate steps of the Nested Sampling algorithm: (a) On entry we have $n$ objects with $L>L^*$ or $\chi <\chi^*$. (b) The one with largest $\chi$ and call it $\chi^*$, removing it from our list which now contains $n-1$ objects sampled from the Prior. (c) We generate a new object, sampled from the Prior once again, but this time constrained to lie within the new Likelihood domain.}
  \label{fig:select}
\end{figure}

Nested sampling (Skilling, 2006) is a modern Bayesian technique which transforms the multidimensional Evidence integral of Equation (\ref{equ:Evidence}) into a simpler one dimensional one. This is done by considering the prior mass $\chi$ and its constituent elements $d\chi = P(\theta|\mathbb{I})d\theta$. These can be summed up as follows:
\begin{equation}
\displaystyle
\chi(\lambda)=\int_{L(\mathbb{D}|\theta,\mathbb{I})>\lambda}P(\theta|\mathbb{I})d\theta
\end{equation}
This covers all the prior mass corresponding to a parameter space with likelihood greater than $\lambda$. This provides for the required transformation to one dimension. The Evidence integral then becomes:
\begin{equation}
\displaystyle
\mbox{Evidence}=\int_0^1 L(\chi) \mbox{d}\chi
\end{equation}
Because of the way we have ordered our elements in terms of Likelihood we can evaluate $L_i=L(\chi_i)$ at a sequence of $m$ points in the parameter space having decreasing Likelihood as shown in Figure \ref{fig:contours}.
\begin{equation}
0<\chi_m<\chi_{m-1}<...<\chi_2<\chi_1<1
\end{equation} 
If we set $h_i=\chi_i-\chi_{i-1}$ the above ordering allows us to evaluate the integral as a weighted sum of the Likelihood:
\begin{equation}
\label{equ:t}
\displaystyle
\mbox{Evidence}=\sum_{i=1}^m{h_iL_i}
\end{equation}
Since we expect that the largest contributions to the integral to come from relatively small regions of peaked Posterior it makes more sense to sample points in $\chi$ at a logarithmic rate instead of a linear one so that initial sampling of the shallow Posterior is rapid. We therefore take:
\begin{equation}
\chi_1=t_1,\quad\chi_2=t_1t_2,\quad...,\quad\chi_i=t_1t_2...t_i
\end{equation}
where each of the $t_i$, known as the \emph{shrinkage ratio}, lies between 0 and 1. In practice the Nested Sampling algorithm implements these concepts as follows:
\begin{enumerate}
\item $N$ objects are sampled randomly from within the prior and their likelihoods are evaluated. Initially we have the full prior range from $0$ to $\chi_0=1$ available to sample from.
\item We then select the point with the lowest likelihood ($L_0$) and remove it from the set of samples. The prior volume is then shrunk to  $\chi_1$ with the shrinkage ratio $t$ determining the volume decrease. The value of $t$ is the expectation value of the largest of N random numbers from uniform distribution between 0 and 1, which is given by $N/(N + 1)$.
\item The removed point is replaced with a new one satisfying the hard constraint likelihood $L>L_0$
\item We then increment the Evidence by $L_0\times(\chi_{1}-\chi_{0})$.
\item We iterate over these steps until we satisfy some stopping condition.
\end{enumerate}
The selection of the least likelihood object is illustrated in Fig. \ref{fig:select}.
The method thus works its way up the likelihood surface through nested surfaces of equal likelihood. To terminate the process we can use two conditions. The first is given in (Skilling, 2006) as the number of iterations $k$ required for samples to converge to posterior peaks:
\begin{equation}
\label{equ:stopping}
k-NH>>0
\end{equation}
Secondly we also requires that successive changes in evidence are sufficiently small.

\section{Variational Bayes}

Variational Bayesian methods provide another approach to parameter estimation and model selection. The basic concept behind this class of methods is to try and approximate the Posterior distribution with a simpler probability distribution. One can then optimize the parameters in this approximation so that it is as close as possible to the true Posterior. It turns out that such an optimal form for the approximating distribution does indeed exist and this is usually very easy to deal with analytically. There is then no need to sample from the Posterior as important quantities such as the mean can be derived analytically from the approximation. Similarly the Evidence can be worked using standard integration techniques on the same approximation. Variational Bayesian methods thus reduce the computationally complex sampling and integration problems to a relatively easy optimization one.

\begin{figure}
  \centering
    \includegraphics[width=2.5in]{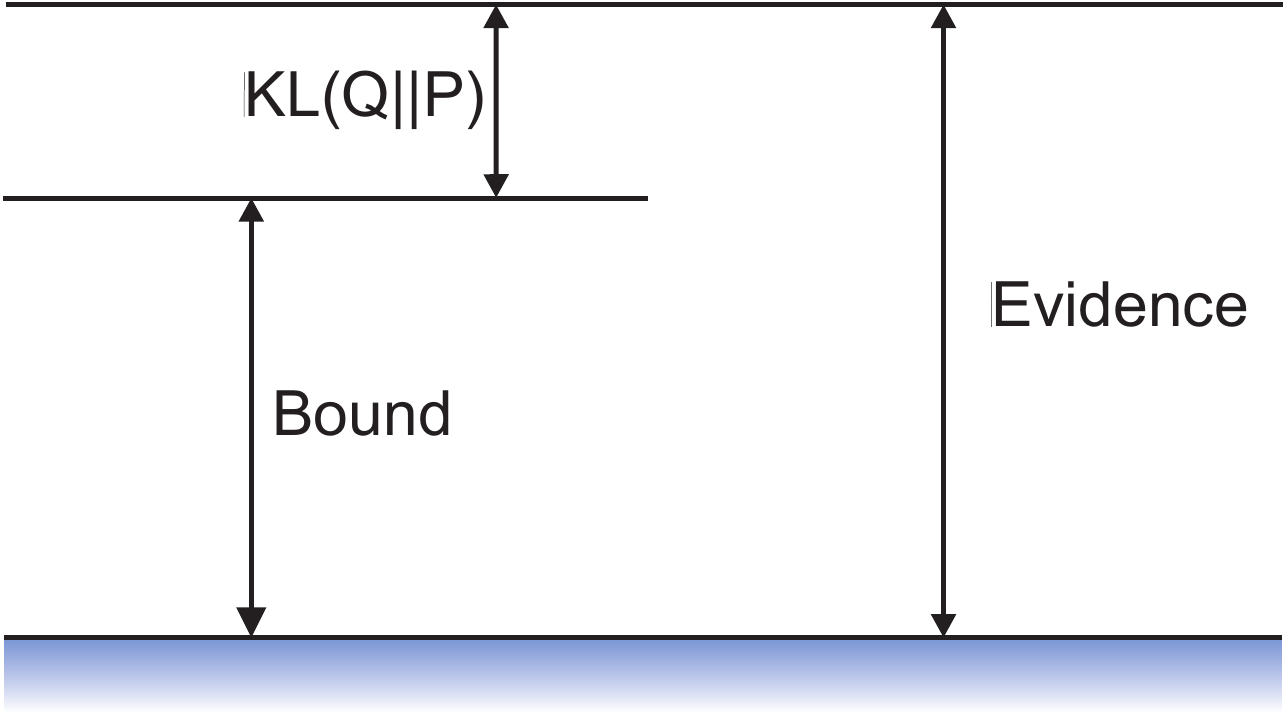}
  \caption[Kullback-Leibler Diveregence]{A diagram based on that in (Zarb Adami, 2003) showing the relationship between the KL divergence and the actual Evidence. Here the value of the Bound is given by the integral given in Equation (\ref{CKL2}).}
  \label{fig:CKL}
\end{figure}

Though there are many optimization algorithms in the literature which are used to find the best approximation the most widely used one is the Expectation-Maximization (EM) algorithm (Zarb Adami, 2003). We give a brief overview of the method here, a more complete explanation can be found in (Mackay, 2003). The sensible question to ask when approximating a complex function by some simpler one $Q(\theta)$ is the amount of information lost in this process. A suitable metric describing this quantity which one can use to quantify the ``distance'' between the original distribution and the approximation is the Kullback-Leibler (KL) divergence (Kullback, 1959), denoted by $D_{\mbox{\scriptsize{KL}}}$. This is given by:
\begin{equation}
D_{\mbox{\scriptsize{KL}}}(Q||P)=\int{Q(\theta)\log\left(\frac{Q(\theta)}{P(\theta|\mathbb{D},\mathbb{I})}\right)}d\theta
\end{equation}
We note that the above expression always returns a non-negative value and, as one would hope, vanishes when $Q(\theta)=P(\theta|\mathbb{D},\mathbb{I})$. Bayes theorem then tells us that:
\begin{equation}
D_{\mbox{\scriptsize{KL}}}(Q||P)=\int{Q(\theta)\log\left(\frac{Q(\theta)P(\mathbb{D}|\mathbb{I})}{L(\mathbb{D}|\theta,\mathbb{I})P(\theta|\mathbb{D,I})}\right)}d\theta
\end{equation}
The Evidence $P(\mathbb{D}|\mathbb{I})$ is a quantity which is independent of the parameters $\theta$ and can thus be taken out of the integral. 
\begin{eqnarray}
D_{\mbox{\scriptsize{KL}}}(Q||P)&=&\int{Q(\theta)\log\left(\frac{Q(\theta)}{L(\mathbb{D}|\theta,\mathbb{I})\pi(\theta|\mathbb{I})}\right)d\theta} \\
&&+\log P(\mathbb{D}|\mathbb{I}) 
\end{eqnarray}
The above equation suggests that one could define a cost function $C_{\mbox{\scriptsize{KL}}}(Q||P)$ which we can then seek to optimize by minimizing the Kullback-Liebler divergence $D_{\mbox{\scriptsize{KL}}}$:
\begin{eqnarray}
\label{CKL}
C_{\mbox{\scriptsize{KL}}}(Q||P)&=&D_{\mbox{\scriptsize{KL}}}(Q||P)-\log P(\mathbb{D}|\mathbb{I}) \\
\label{CKL2}
&=& \int{Q(\theta)\log\left(\frac{Q(\theta)}{L(\mathbb{D}|\theta,\mathbb{I})\pi(\theta|\mathbb{I})}\right)d\theta} 
\end{eqnarray}
Since $D_{\mbox{\scriptsize{KL}}}(Q||P)\geq0$ the following inequality always holds:
\begin{eqnarray}
C_{\mbox{\scriptsize{KL}}}(Q||P) &\geq & -\log P(\mathbb{D}|\mathbb{I})\\
\Rightarrow-C_{\mbox{\scriptsize{KL}}}(Q||P) &\leq& \log P(\mathbb{D}|\mathbb{I})
\end{eqnarray}
We thus see how minimizing the cost function is equivalent to maximizing a lower bound on the Evidence value. Figure \ref{fig:CKL} shows the relation between the lower bound to the Evidence and the Kullback-Leibler divergence.
Fortunately enough, in most cases the lower bound to the Evidence is tight enough to be used in place of the true Evidence in model selection procedures (Miskin, 2000).  After having optimized the approximating distribution one can then perform inference with the approximation in place of the real Posterior.

\section{Comparisons}
In this section we can compare Nested Sampling and Variational Bayes as applied to two engineered problems, the first being Polynomial Selection and the second being the fitting of Gaussian Mixture Models. We use MATLAB in this preliminary study to speed up the coding process. Naturally both methods are coded in the same language to try and ensure a fairer comparison of the computational times.

\subsection{Polynomial Selection}

\begin{figure}
\begin{center}
$\begin{array}{c@{\hspace{0.05in}}c}
\includegraphics[width=1.6in]{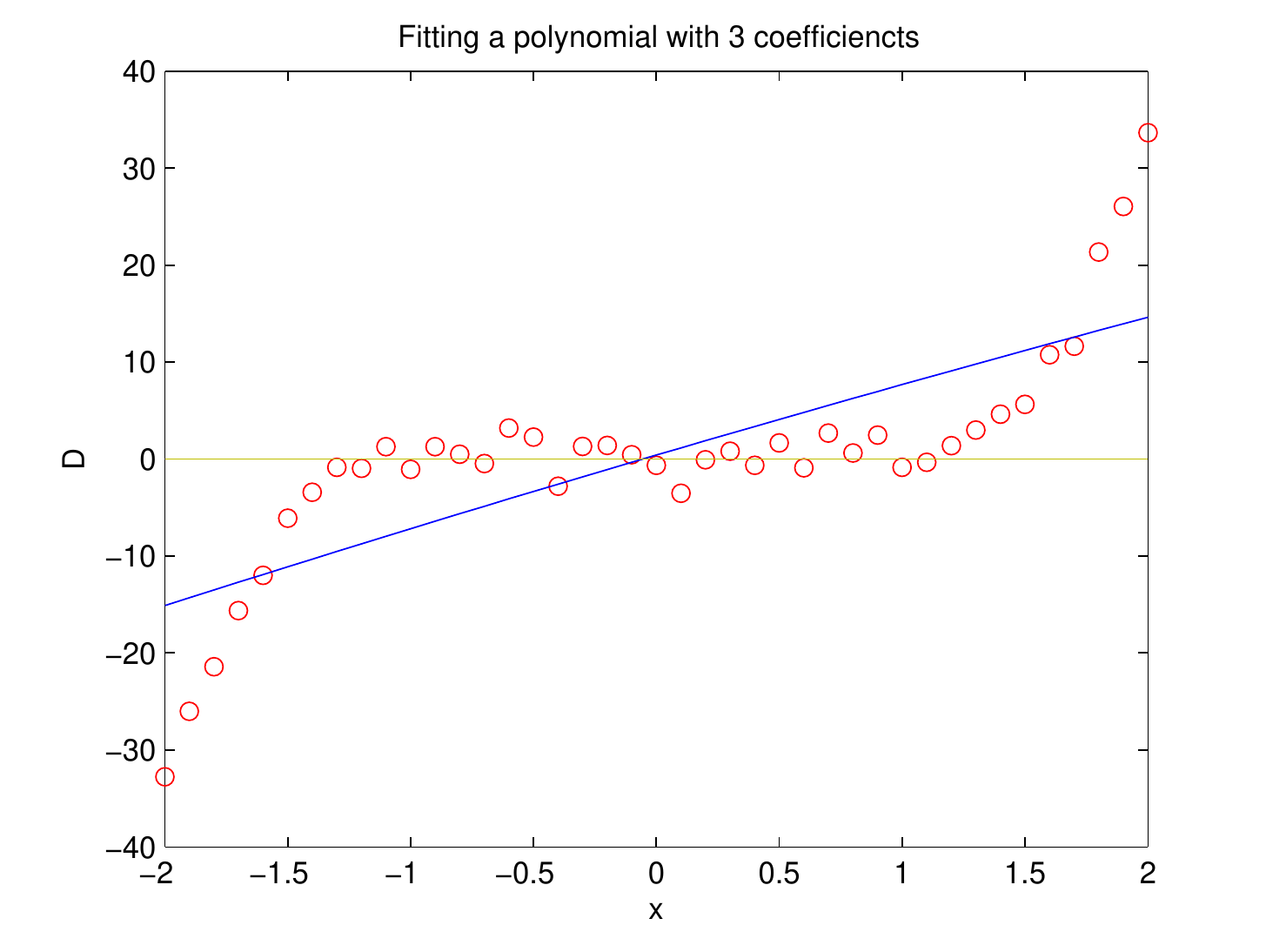}  &  
\includegraphics[width=1.6in]{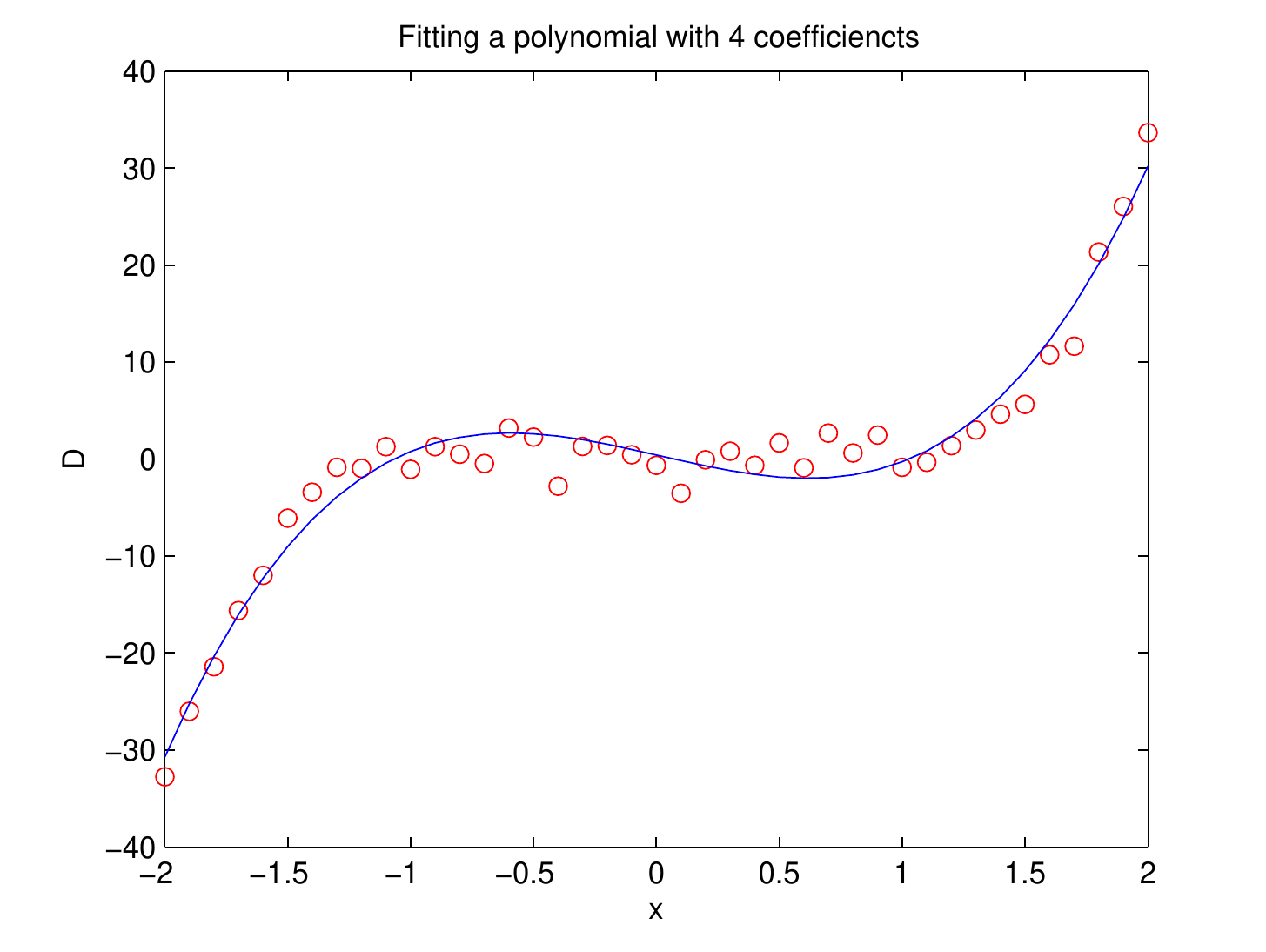} \\
\includegraphics[width=1.6in]{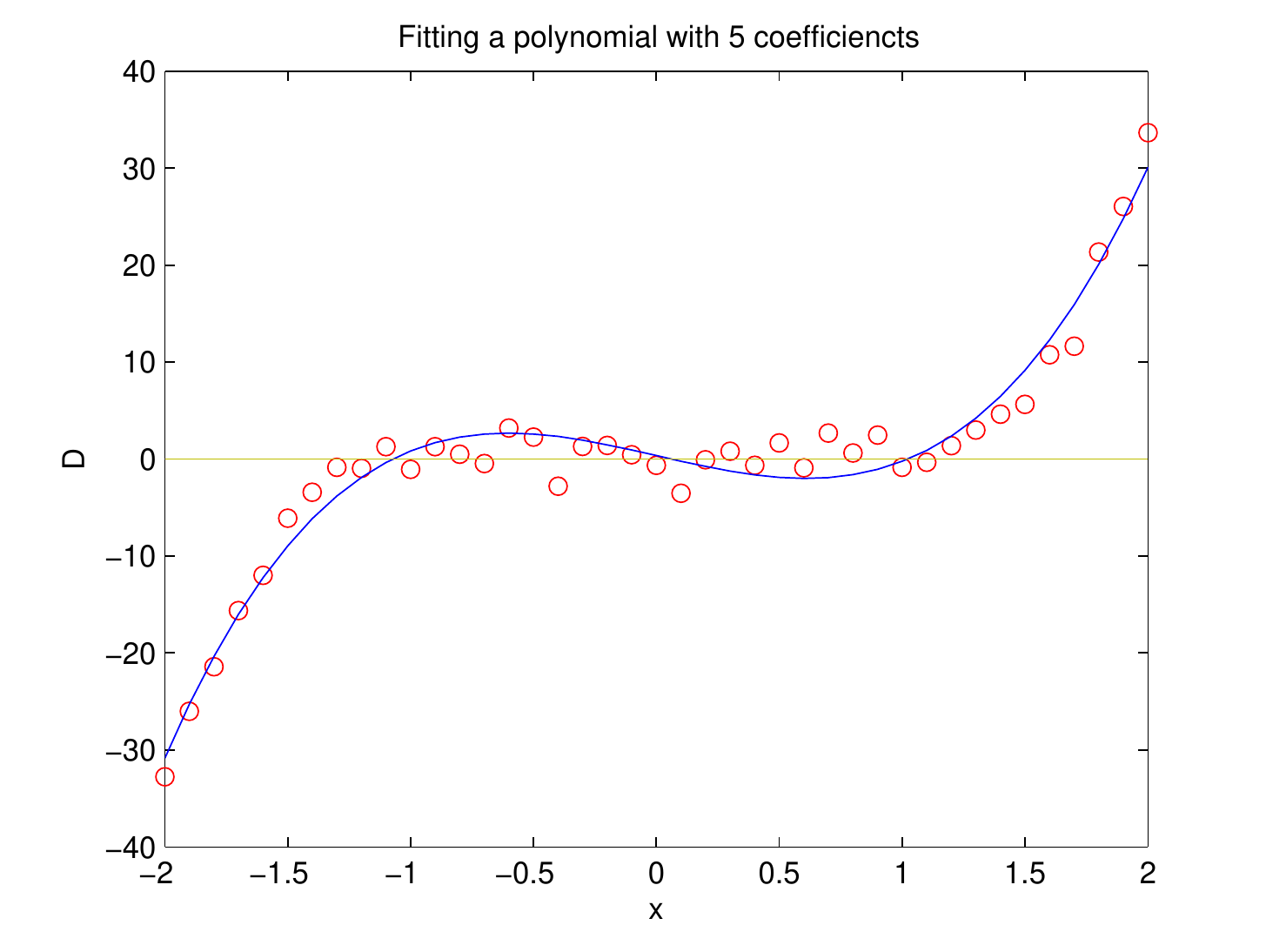}   &
\includegraphics[width=1.6in]{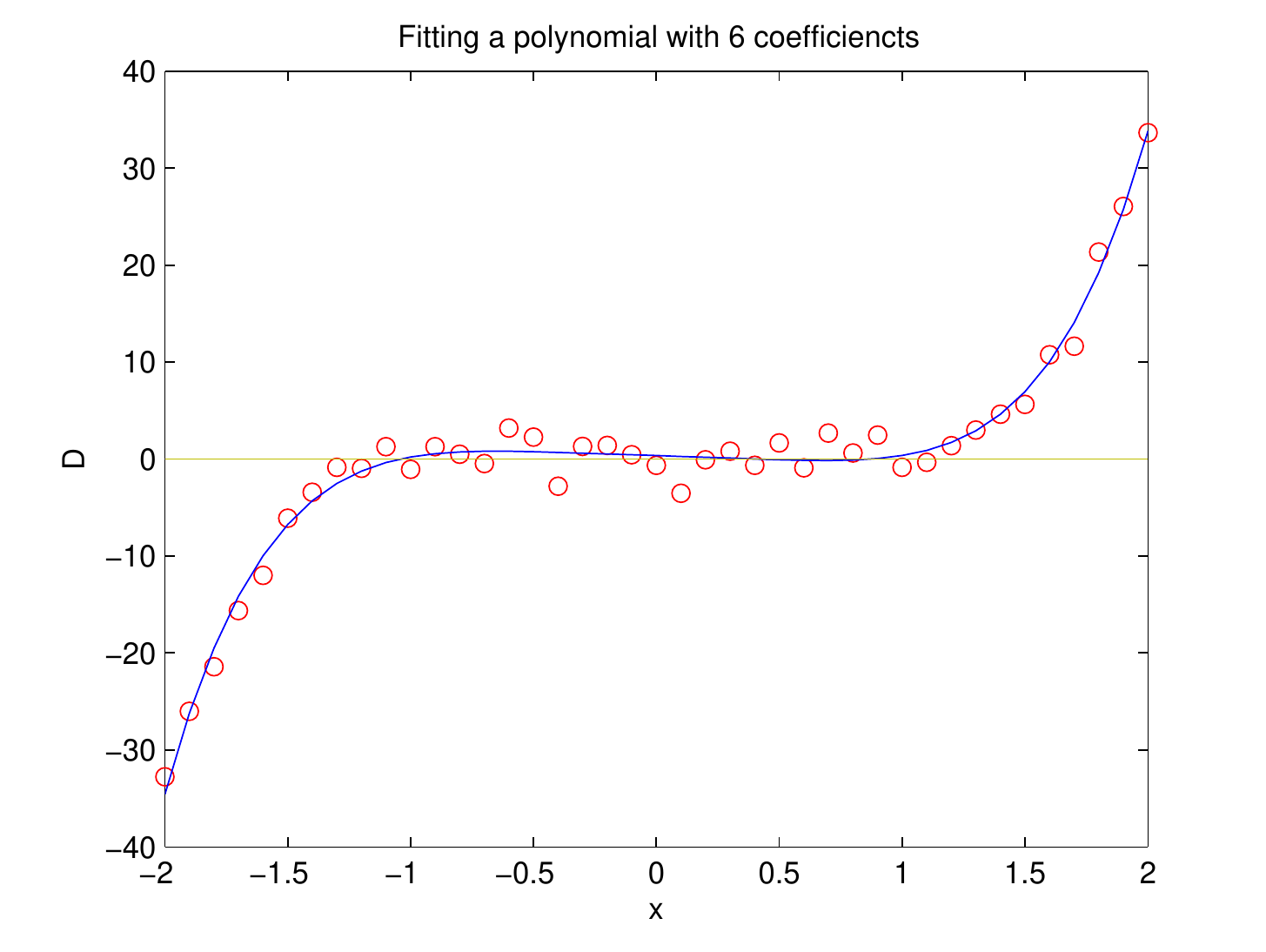}   \\
\end{array}$
\end{center}
\caption[Polynomial Fitting using VB]{Plots showing the data set fitted with polynomials of different orders using Variational Bayes.}
\label{fig:polyVBFitted}
\end{figure}

When one has a sparse data set it is often useful to infer a smooth curve that best fits with the observed points. This can facilitate further studies of the data as procedures like integration or the finding of extrema then become very easy. Of course one might also want to fit data which has been generated by a physical law which has polynomial dependence on some parameter. The task at hand is thus to select the degree of the polynomial to use as well as calculate its coefficients. If a low order polynomial is chosen then it might be difficult to fit the data well whilst if the polynomial order is too high then there is a risk of over-fitting. We denote our data set $\mathbb{D}$, where $\mathbb{D}$ contains a set of points $\left\{(x_i,D_i)\right\}$. Here $1\leq i \leq I$ and $I$ is the total number of data points. The $x_i$ are the abscissa values whilst the $D_i$ are the ordinate values. In the general case one can then construct an interpolation model using a set of $K$ fixed basis functions. We thus have:
\begin{equation}
\label{equ:dataPoly1}
\displaystyle
D_i = \sum_{n=1}^N w_n f_{ni} + \epsilon_i
\end{equation}
Here $\mathbf{f}=f_{ni}$ is a matrix constructed using the interpolating basis functions evaluated at the abscissa values $x_i$. If we decide to choose polynomials then we get $f_{ni}=F_n(x_i)$ where $F_n(x)=x^{n-1}$. The value of $\epsilon_i$ in Equation (\ref{equ:dataPoly1}) represents the noise affecting each measurement, which we can assume to be Gaussian with inverse variance $\gamma$. If we assume that the Gaussian noise is independent then the Likelihood function becomes:
\begin{equation}
\displaystyle
\label{equ:polyLike}
L(\mathbb{D}|\bm{w},\gamma, \mathbb{I})=\prod_{i=1}^{I}{G\left(D_i | \sum_{n=1}^{N}w_nf_{ni}, \gamma \right)}
\end{equation}
We generate the data for this study by computing the actual values of a polynomial within a predefined interval $\left[a,b\right]$ and adding the Gaussian noise of inverse variance $\gamma$ using MATLAB's inbuilt functions. Unless otherwise stated all the plots shown in this section are based on a data set containing 40 data points generated in the interval $\left[-2,2\right]$ from a quintic curve with added noise having $\sigma=\gamma^{-1/2}=2$.
We first tackle the situation using Variational Bayesian and then move on to use Nested Sampling. For the variational solution we can assign conjugate priors:
\begin{eqnarray}
P\left(\mathbf{w} | \mathbb{I}\right) & = & G\left(\mathbf{w}| 0, a\bm{^w} \right) \\
P\left(\gamma | \mathbb{I}\right) & = & \mbox{Gamma}\left(\gamma | a^\gamma, b^\gamma\right)
\end{eqnarray}
We can set the three arbitrary parameters $a^{\bm{w}},a^\gamma$ and $b^\gamma$ accordingly to modify the shape and scale of the priors. These are chosen to make the priors as broad as possible, to simulate our ignorance of the underlying phenomena. The update equations, given in (Zarb Adami, 2003) and (Miskin, 2000), are shown below:
\begin{eqnarray}
\displaystyle
\mathbf{\tilde{w}}&=&a^{(w)}\mathbf{I}+\left\langle \gamma\right\rangle_Q\mathbf{ff}^T \\
\mathbf{\hat{w}}&=&\mathbf{\tilde{w}}^{-1}\left\langle \gamma\mathbf{f}\right\rangle_Q \\
\bar{a}^{(\gamma)}&=&a^{(\gamma)}+\frac{1}{2}\sum_{i=1}^I\left\langle \left(D_i-\sum_{n=1}^Nw_nf_{ni}\right)^2\right\rangle_Q \\
\bar{b}^{(\gamma)}&=&b^{(\gamma)}+\frac{I}{2}
\end{eqnarray}
The iteration of these equations leads to the values for the optimal parameter set describing the data. For our 40 point data set we can try fitting different order polynomials. The results are shown in Figure \ref{fig:polyVBFitted}.
One can observe how inefficient the fitting is when we use a quadratic and how this improves once we use a cubic curve. Fitting with a quintic ($n=6$) we obtain:
\begin{eqnarray}
\mathbf{w} &=& (0.35, -0.82, -0.03, -0.29, -0.04, 1.19)\\
\sigma & = & 1.65
\end{eqnarray}
These results are not too far from the true values of $\mathbf{w} = (0, 0, 0, 0, 0, 1)$ and $\sigma=2$ and an element of inconsistency between the two is acceptable as we have taken a large noise factor and not so many data points (40). Note that the odd components in the weight vector, which correspond to the even polynomials $x^{2n}$, are especially close to zero. This is because the parity of these components (even) is different from the parity of the mechanism generating the data (odd). Hence these contributions are fitted to zero as they can offer no explanation of the data.
\begin{figure}
\begin{center}
$\begin{array}{c@{\hspace{0.05in}}c}
\includegraphics[width=1.6in]{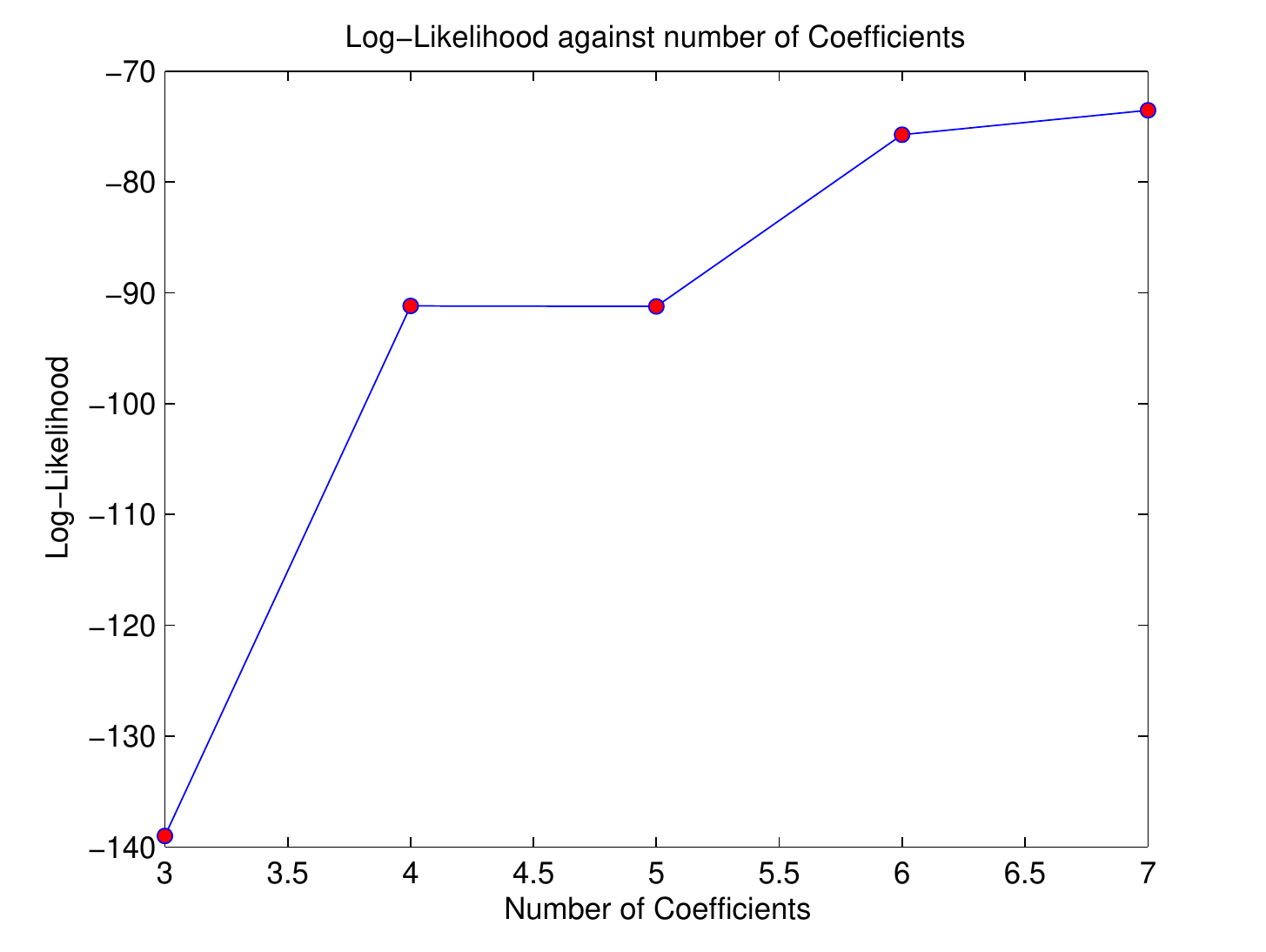}  &  
\includegraphics[width=1.6in]{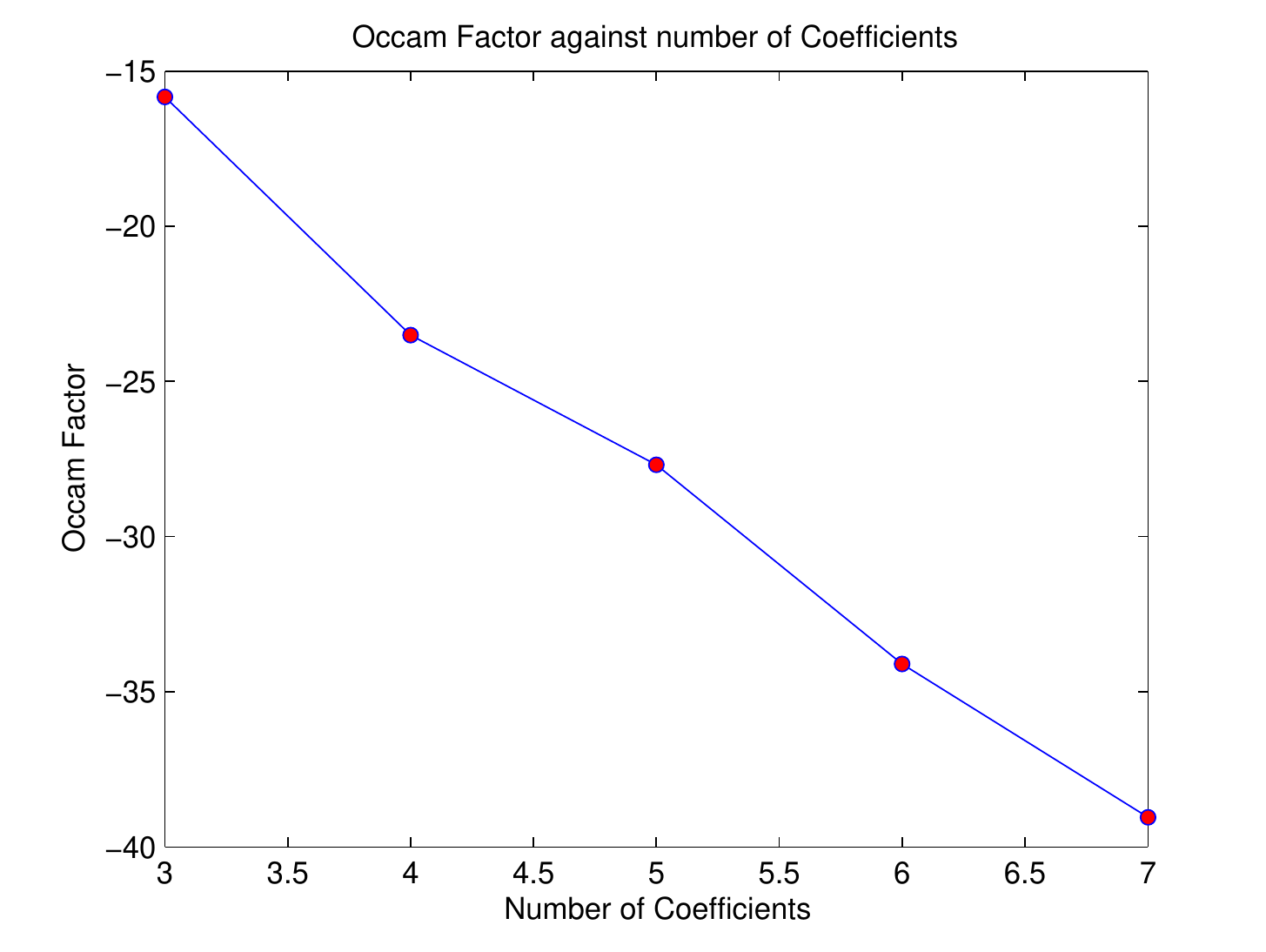} \\
\end{array}$
$\begin{array}{c}
\includegraphics[width=1.6in]{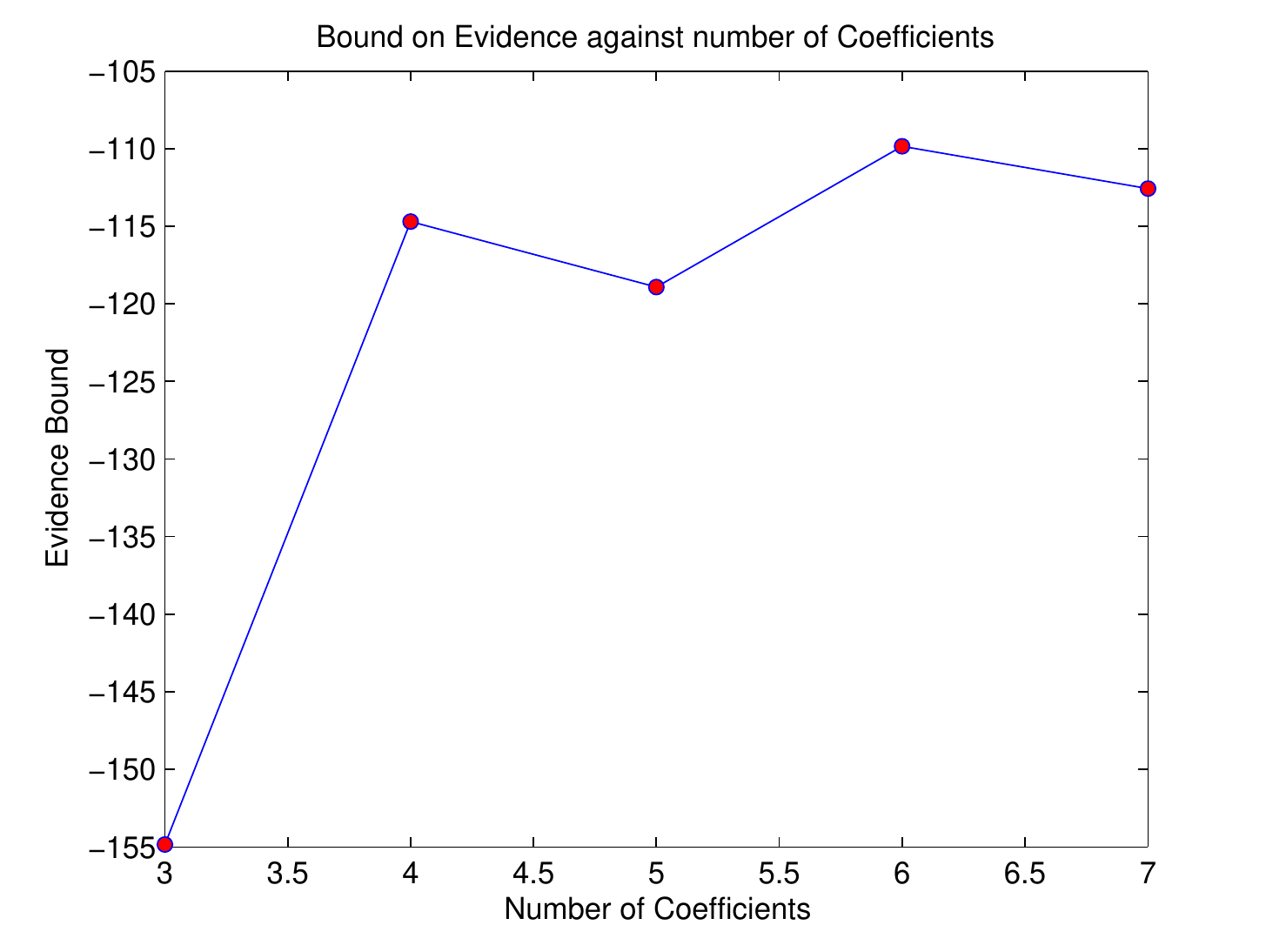}   \\
\end{array}$
\end{center}
\caption[Polynomial Selection using VB]{Polynomial Selection using VB. One can see the Log-Likelihood $L$ (top left), the Occam factor (top right) and the Evidence (bottom).}
\label{fig:polyVB}
\end{figure}
Referring to Figure \ref{fig:polyVB} we notice that as expected the Likelihood always increases with the number of parameters used to describe the data. However the Evidence is also dependent on an Occam factor which penalizes more complex models and is thus peaked, beginning to decrease past $n=6$. The peak at $n=6$,  correctly implies that the most probable hypothesis is that the data was generated using a 6 parameter polynomial which is a quintic. 

To implement the Nested Sampling algorithm for this example we take Skilling's code (Skilling, 2004) as a skeleton and modify the Likelihood and exploration functions appropriately to represent Equation (\ref{equ:polyLike}) . We set uniform priors for $\gamma$ and $\bm{w}$ accordingly. 
Exploration of the parameter space using a different variable step size for each separate parameter. The algorithm was then run with 36 objects. Nested Sampling is much slower than the Variational method as the average time of computation was in the region of 300 seconds, an order of magnitude slower then Variational Bayes which takes around 20 seconds for a typical run. This difference in speed is most probably due to the fact that Nested Sampling is an MCMC method, involving the probabilistic exploration of a parameter space whilst Variational Bayes is analytical and has no such probabilistic dependence.
\begin{figure}
\begin{center}
$\begin{array}{c@{\hspace{0.05in}}c}
\includegraphics[width=1.6in]{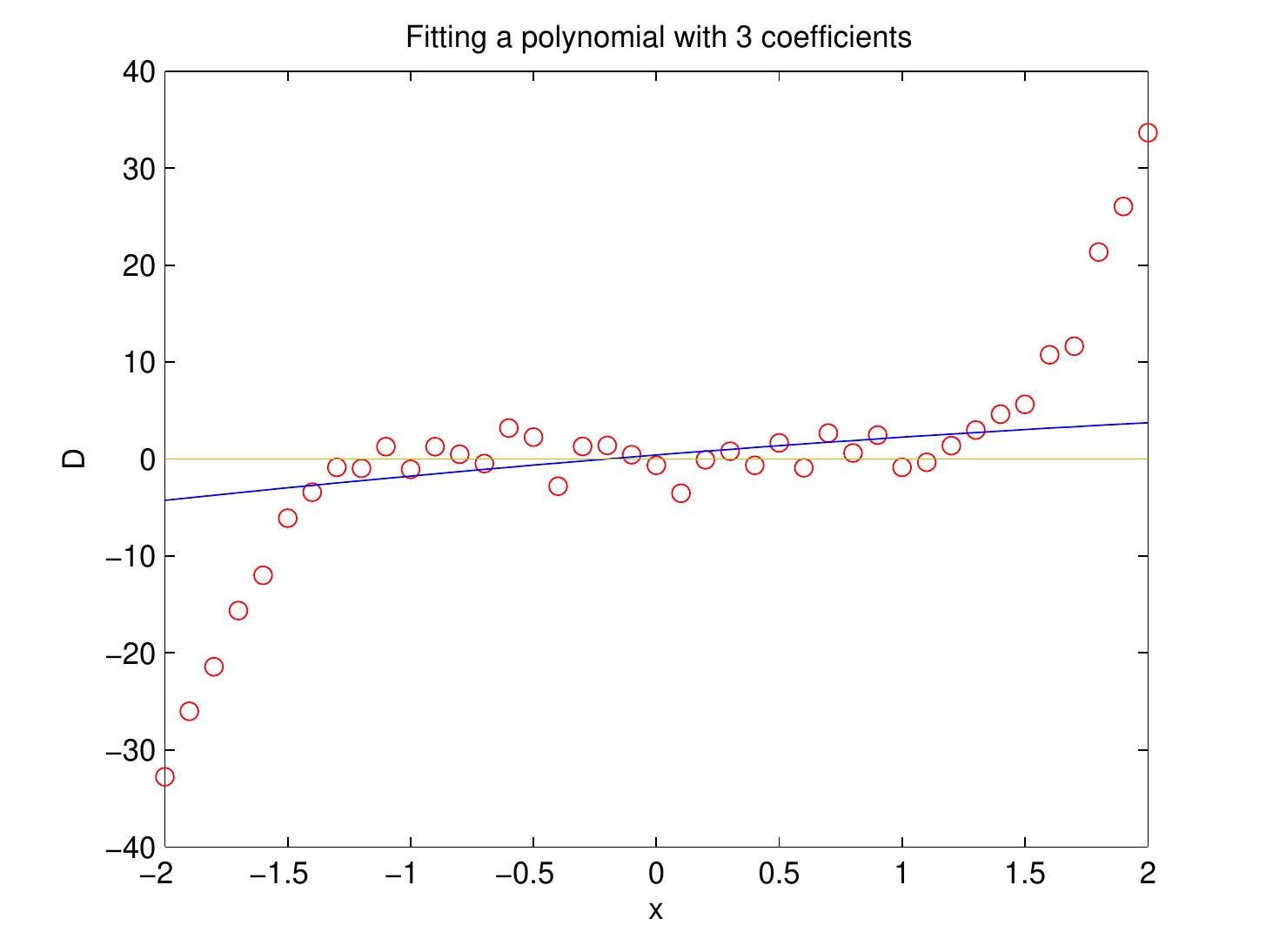}  &  
\includegraphics[width=1.6in]{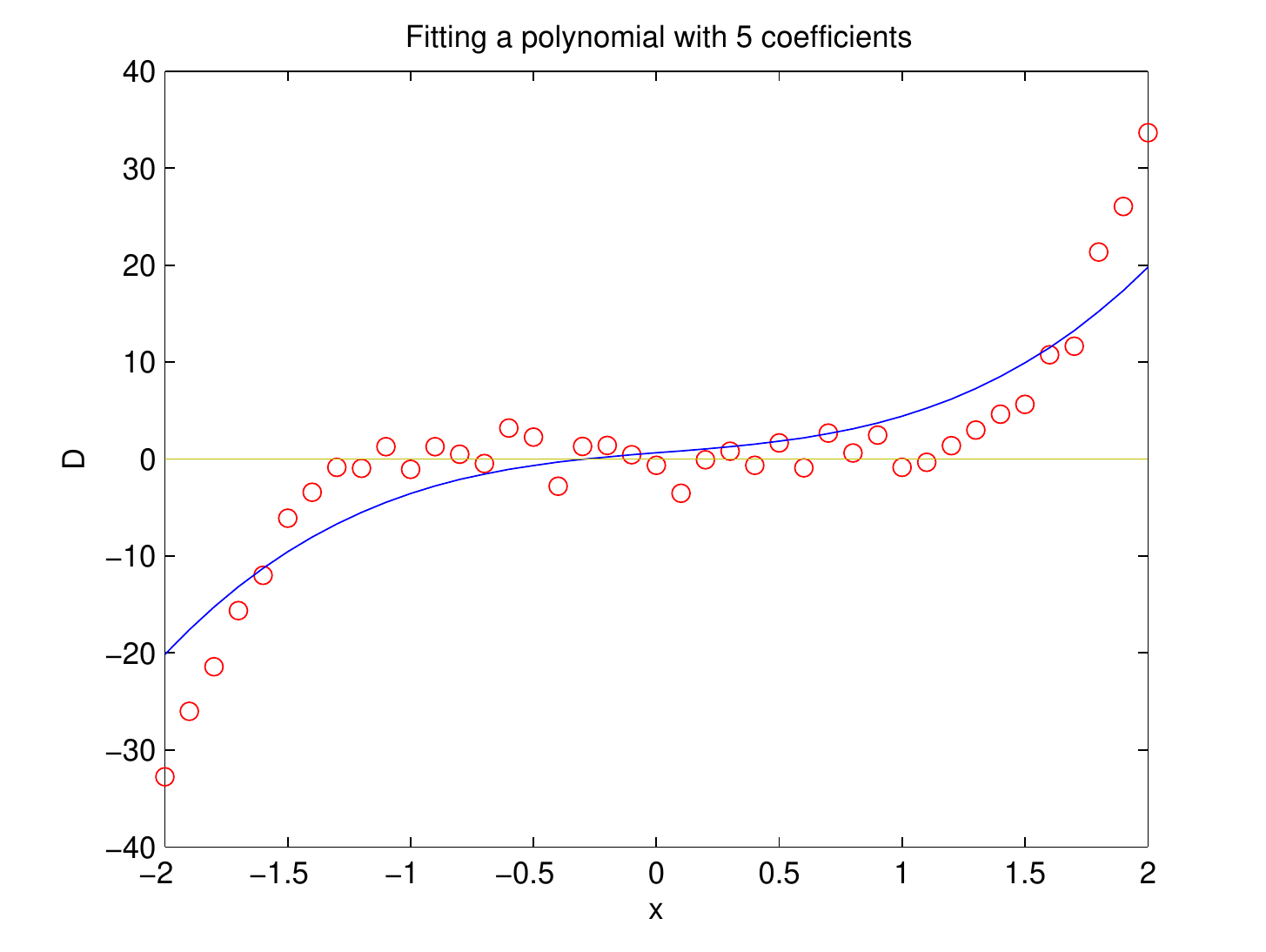} \\
\includegraphics[width=1.6in]{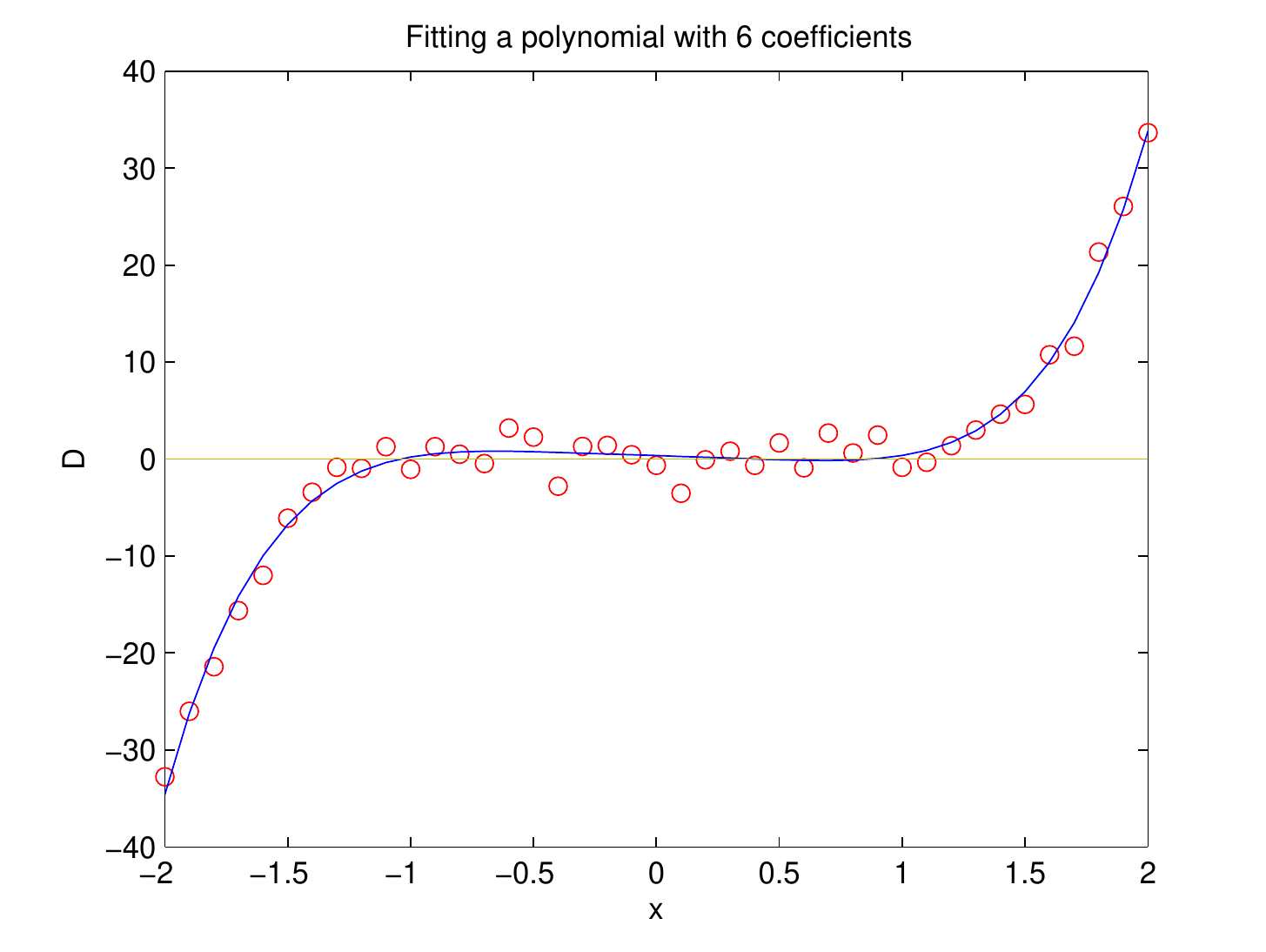}   &
\includegraphics[width=1.6in]{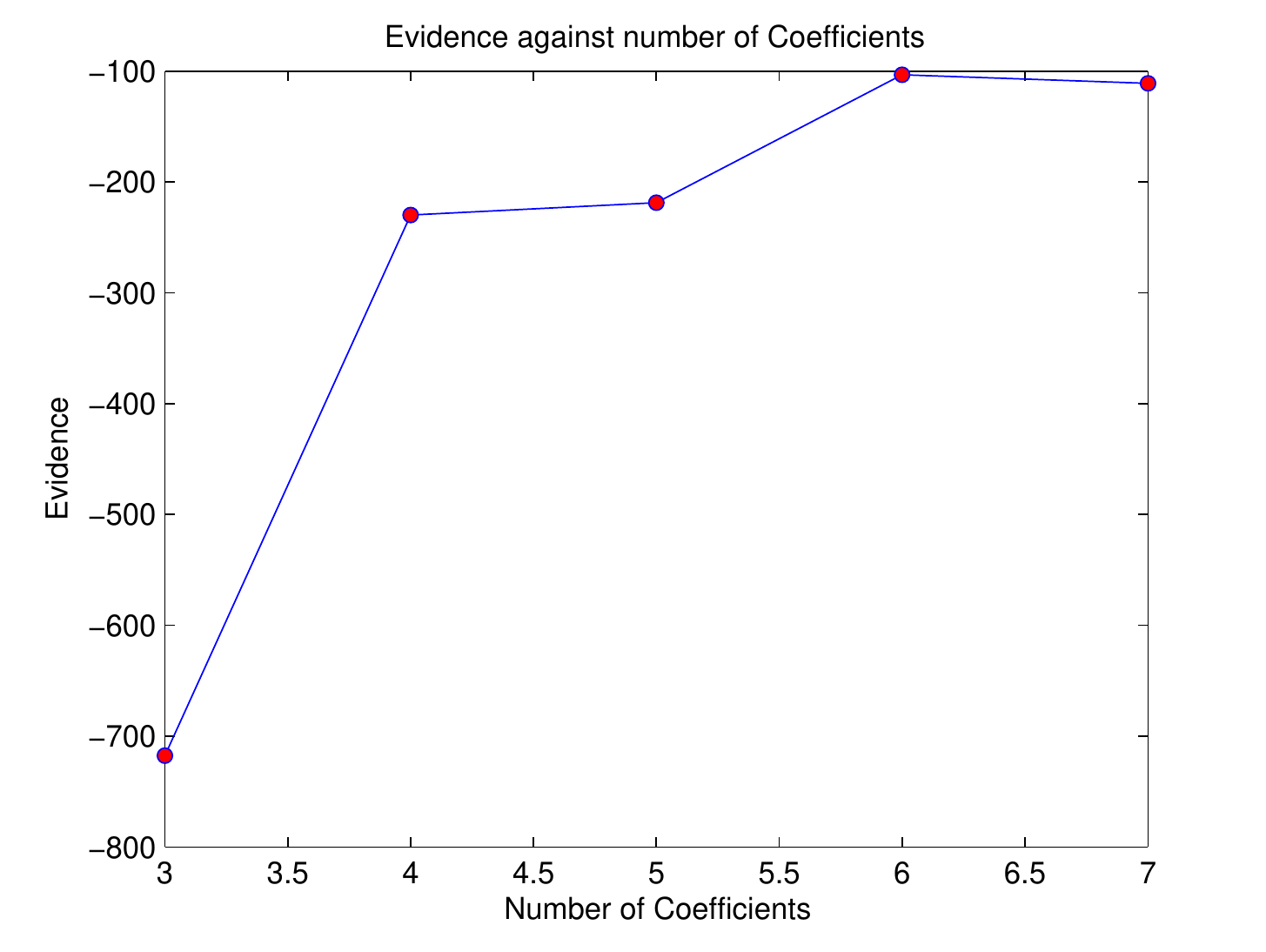}   \\
\end{array}$
\end{center}
\caption[Polynomial Selection using Nested Sampling]{Polynomial Selection using Nested Sampling. The bottom right shows the Evidence whilst the others are different polynomial fits.}
\label{fig:polyNested}
\end{figure}
Plots showing the Evidence and fits are given in Figure \ref{fig:polyNested}. Note that within the Nested Sampling framework we have no explicit ``Occam Factor''. Rather the increase in parameters penalizes the complex models by giving the algorithm more space to explore and thus longer time to reach regions of higher Likelihood. This means that when calculating the Evidence the higher Likelihood values are multiplied by smaller weights as shown in Equation (\ref{equ:t}) resulting in a lower Evidence value over all. Again we get the Evidence peaking at the correct value of $n=6$ with the optimum Evidence for both methods at around $\log (\mbox{Evidence})=-110$. The coefficients themselves are also in very good agreement with the ones obtained using Variational Bayes and we get:
\begin{eqnarray}
\mathbf{w} &=& (0.35, -0.82, -0.03, -0.29, -0.04, 1.19)\\
\sigma & = & 1.65
\end{eqnarray}
In fact the greatest variation is in the relatively inconsequential coefficient of $x^2$ and the value of this difference is $18$\%. This is not such an issue as the coefficient is very small anyway and has no major effect on the final plots. The averaged percentage disagreement in all the other parameters is under $1$\% and the percentage disagreement in the coefficient of $x^5$ is only $0.04$\%. This indicates clearly that the discrepancies from the true parameters are due to the random nature of the Gaussian noise and do not arise because of any fault in either of the algorithms. Note that we give all the results for the parameters in Appendix A, along with the percentage difference between the two methods.

\subsection{Gaussian Mixture Models}
 A Mixture Model is basically a distribution built up using a number of simpler distributions having different parameters. They are often used to obtain or substitute more complex distributions. An illustrative example given in (Miskin, 2000) concerns the size of fruit. The probability distribution will obviously depend on the type of fruit being measured. Instead of having a single complex distribution it would make sense to model the size distribution for each fruit type using a Gaussian and then have a \emph{categorical variable} which gives the probability that a fruit is of a given type. Mathematically this can be expressed in terms of the Likelihood function for a single data point $D_i$, where $i$ labels the data point and can take values $1\leq i\leq I$. If we have $S$ categories then we have:
\begin{equation}
\label{equ:mixture1}
\displaystyle
L(D_i|\theta,\mathbb{I})=\sum_{s=1}^{S} P(s_i=s|\pi,\mathbb{I})P(D_i|\theta_s,\mathbb{I})
\end{equation}
Here $s_i$ is an \emph{indicator} variable for each data point which tells us which distribution created the $i^{\mbox{\scriptsize{th}}}$ data point. These are chosen probabilistically such that $P(s_i=s|\mathbf{\pi},\mathbb{I})=\pi_s$. Also, $P(D_i|\theta_s,\mathbb{I})$ is the Likelihood for a given $s_i$ and data point $D_i$. We henceforth consider the particular case when the mixture is composed of Gaussians, known as a \emph{Gaussian Mixture Model} (GMM).  Equation (\ref{equ:mixture1}) then becomes:
\begin{equation}
\displaystyle
L(D_i|\theta,\mathbb{I})=\sum_{s=1}^{S}\pi_s G(D_i|\mu_s,\sigma_s)
\end{equation}
The parameters $\mathbf{\mu}=(\mu_1,\mu_2,...,\mu_S)$, $\mathbf{\sigma}=(\sigma_1,\sigma_2,...,\sigma_S)$ and $\mathbf{\pi}=(\pi_1,\pi_2,...,\pi_S)$ are collectively referred to as $\theta$. 

\begin{figure}
 \centering
  \includegraphics[width=2.5in]{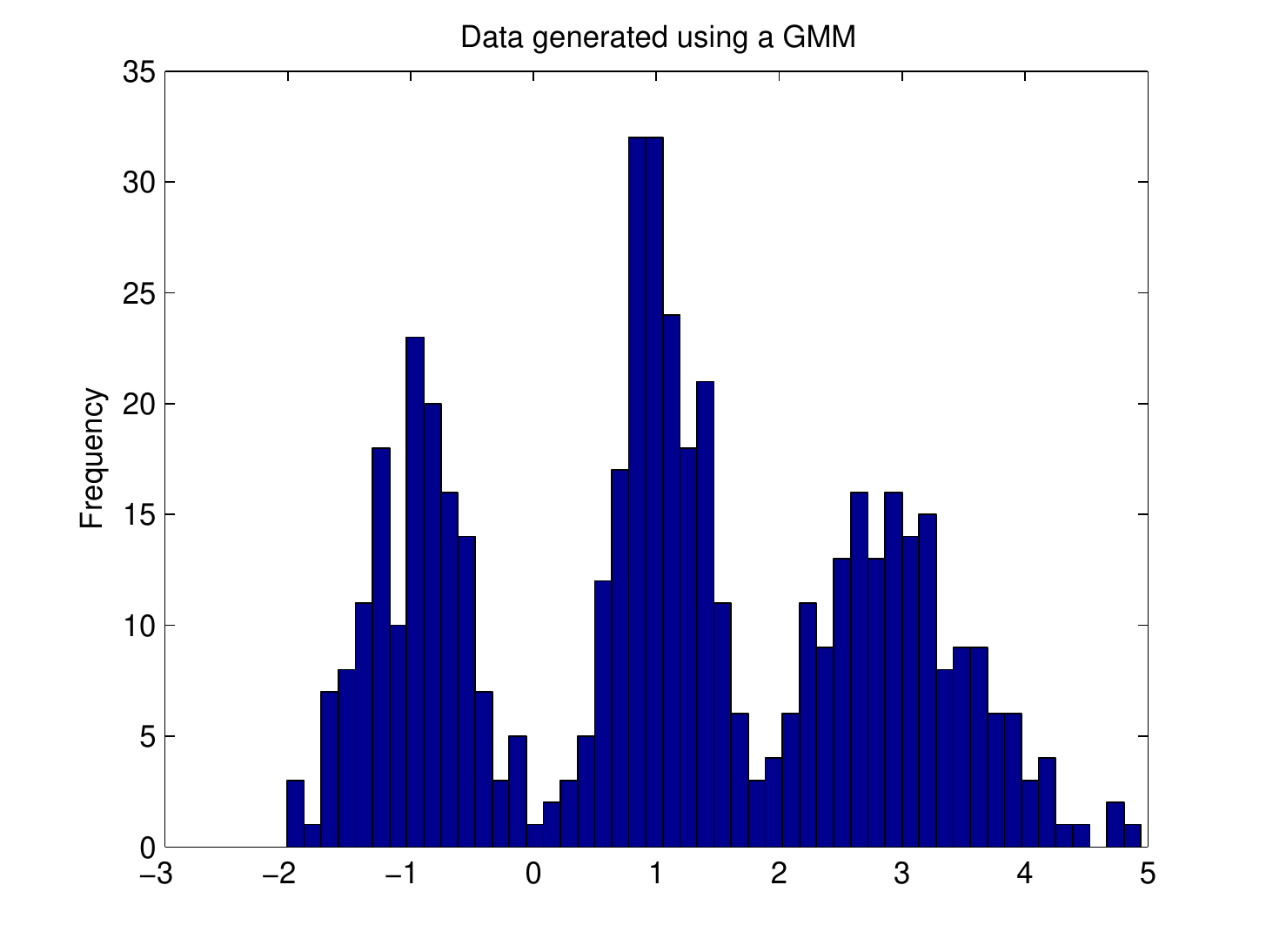}
\caption[Histogram of Gaussian Mixture Model]{Histogram showing the distribution composed of three Gaussians. These had means $\mathbf{\mu}=(-1,1,3)$:, sigma $\mathbf{\sigma}=(0.4,0.3,0.7)$ and were produced in the following ratio $(0.3:0.35:0.35)$. }
  \label{fig:GMM1}
\end{figure}

A histogram of the generated points used in this section is shown in Figure \ref{fig:GMM1}. The data set contains 300 points generated from a Mixture Model composed of three Gaussians. These have means $\mathbf{\mu}=(-1,1,3)$ and sigma $\mathbf{\sigma}=(0.4,0.3,0.7)$ . The points were produced in the following ratio $(0.3:0.35:0.35)$.
The data set is relatively easy to handle for the algorithm because there is only a slight overlap between the different component distributions. 

\subsubsection{Variational Bayesian Methods}
\label{gmmVB}
\begin{figure}
\begin{center}
$\begin{array}{c@{\hspace{0.05in}}c}
\includegraphics[width=1.6in]{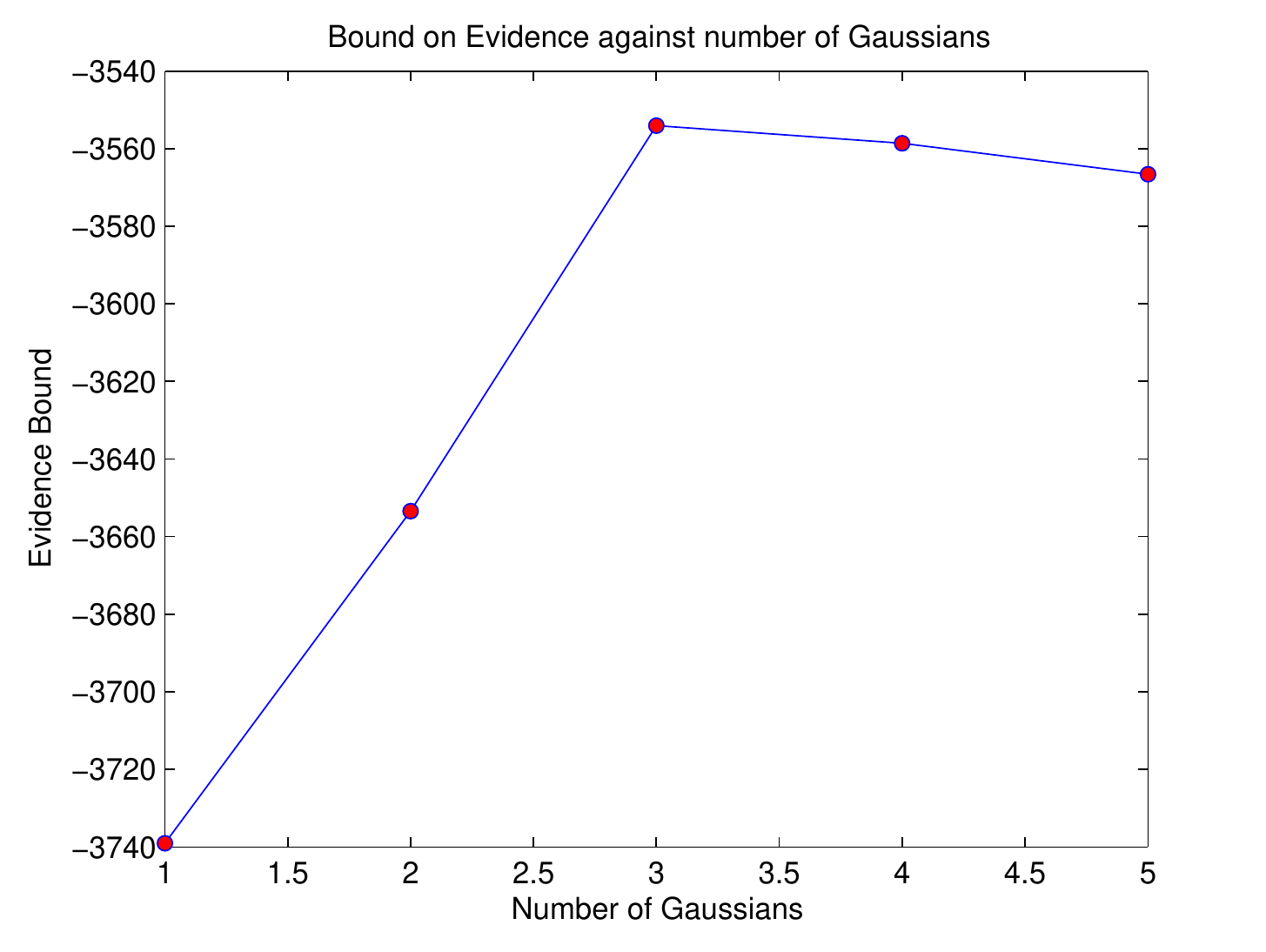}  &
\includegraphics[width=1.6in]{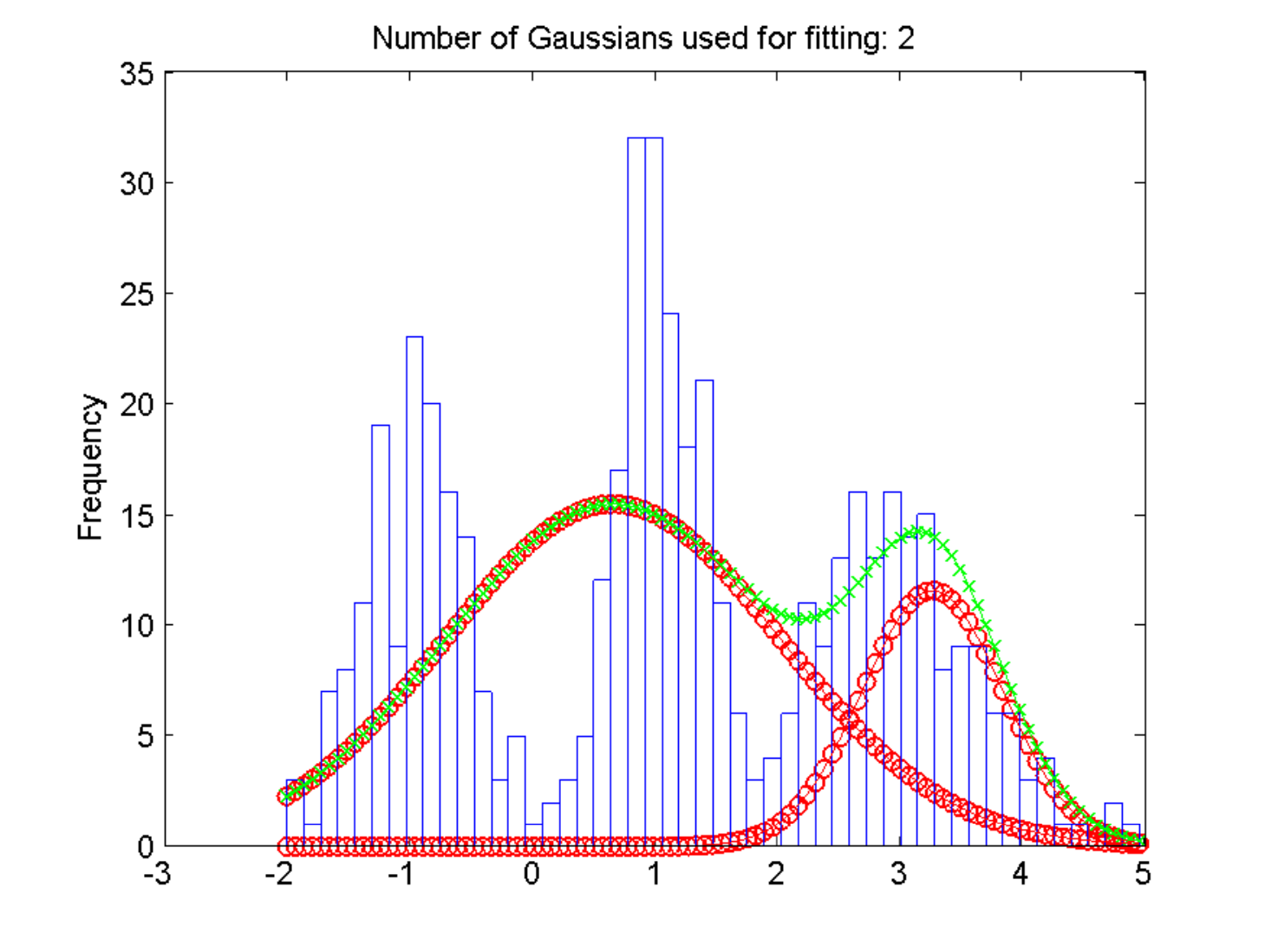} \\
\includegraphics[width=1.6in]{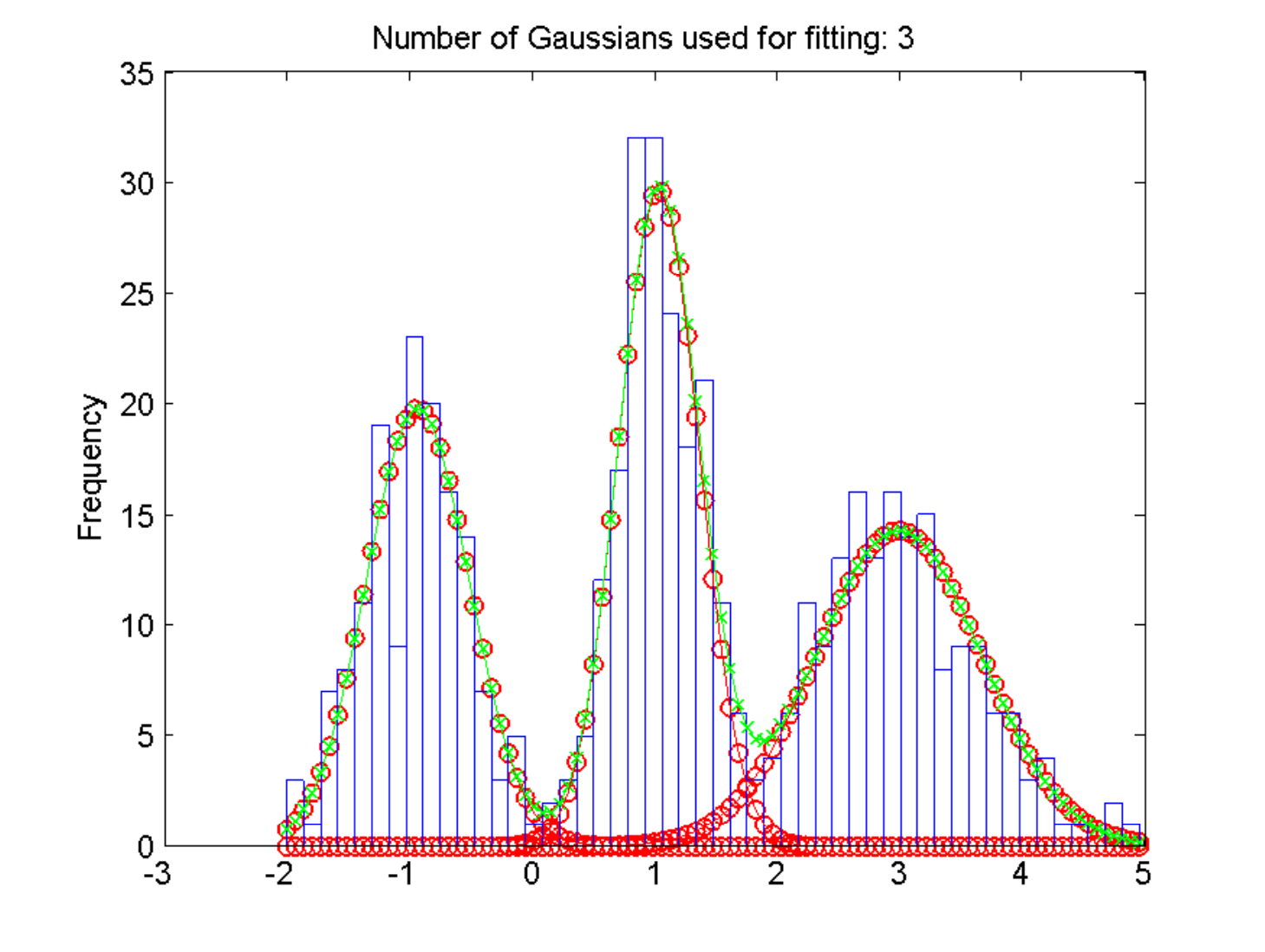}  &  
\includegraphics[width=1.6in]{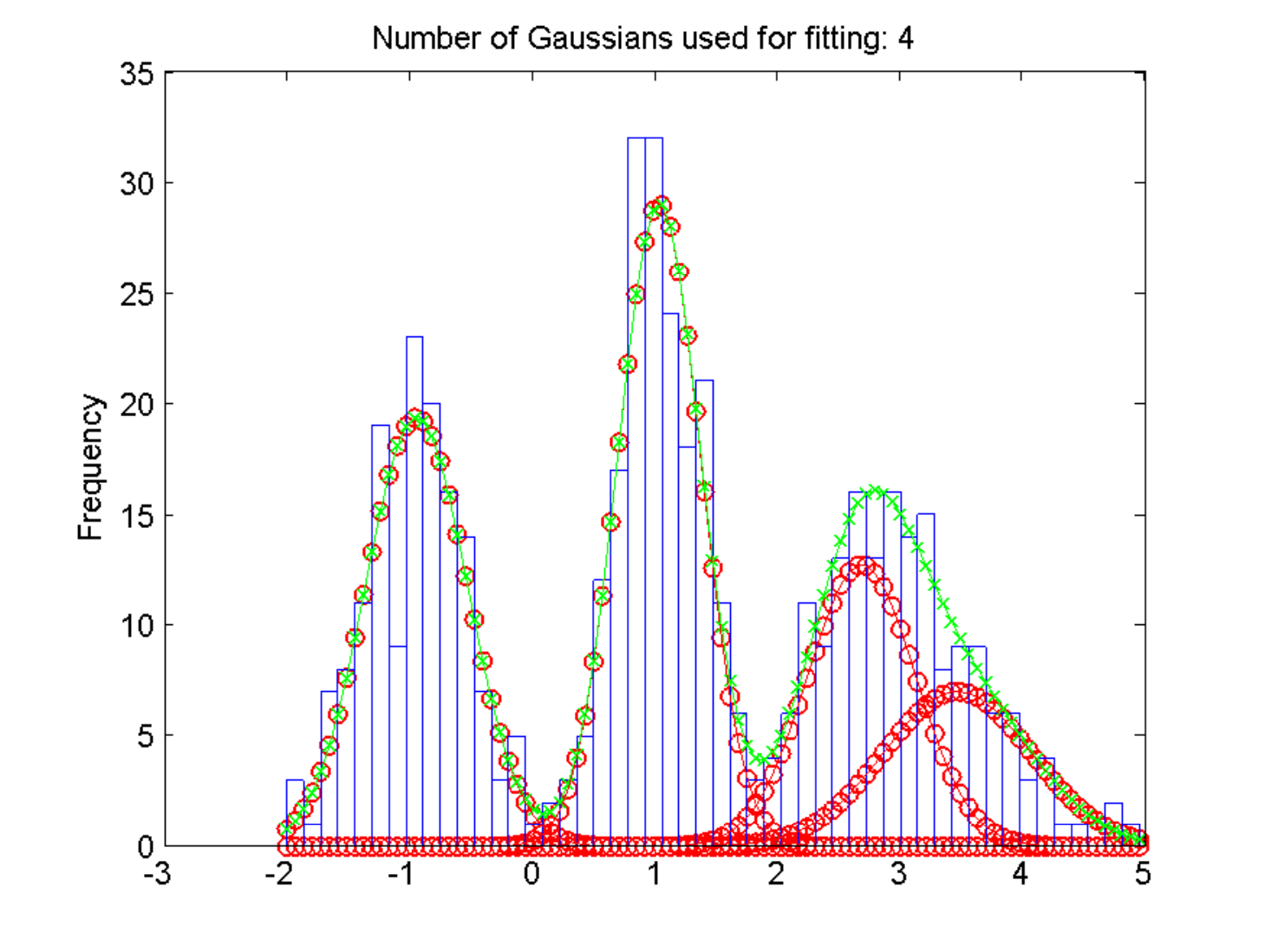} \\
\end{array}$
\end{center}
  \caption[GMM Results using VB]{GMM Results using VB. We can see here three attempts to fit the data with different amounts of Gaussians as well as a plot of the Evidence bound (top left).}
  \label{fig:GMM3}
  \end{figure}

We first set out to attack the problem using Variational Bayesian methods, fitting the data with $S$ Gaussians. We use conjugate Priors as required by the Variational Bayesian algorithm, using $s$ to label the variables pertaining to each of the $S$ Gaussians.
\begin{eqnarray}
P(\mu_s|\mathbb{I})&=&G(\mu_s| m_{0s}, \tau_{0s}) \\
P(\beta_s=\frac{1}{\sigma_s}|\mathbb{I})&=&G(\beta_s| b_{0s}, c_{0s}) \\
P(\pi|\mathbb{I})&=&D(\pi_s| \lambda_{0s}) 
\end{eqnarray}
The first two equations are Gaussian Priors for the parameters pertaining to a single Gaussian in the mixture whilst the third, $D$, represents a Dirichlet distribution over the categorical weights with mixing hyperparameter $\lambda_{0s}$ as described in Appendix A. If we assume the independence of the separate data points and we consider all the distributions forming the mixture, the Prior becomes:
\begin{equation}
\displaystyle
P(\theta|\mathbb{I})=P(\pi|\mathbb{I})\prod_{s=1}^{S}P(\beta_s|\mathbb{I})\prod_{s=1}^{S}P(\mu_s|\mathbb{I})
\end{equation}
Note that here $\beta_s,\mu_s,b_{0s},c_{0s},\pi$ and $\lambda_{0s}$ each refer to a single one of the Gaussian distributions forming the mixture. The joint Likelihood of any single data point is given by Equation (\ref{equ:jointDataInd}). 
\begin{equation}
L(D_i,s_i|\theta)=L(s=s_i|\pi)G(D_i|\beta_s,\mu_s)
\label{equ:jointDataInd}
\end{equation}
If we assume that the data are independent then we construct the final Likelihood by taking the following product:
\begin{equation}
\displaystyle
L(\mathbf{D},\mathbf{s}|\theta)=\prod_i^{I}L(s_i=s)G(D_i|\beta_s,\mu_s)
\label{equ:gmmLikelihood1}
\end{equation} 
Here $\mathbf{D}=(D_1,D_2,...,D_I)$ and $\mathbf{s}=(s_1,s_2,...,s_I)$. The inclusion of the indicator variable introduces additional complexities into the derivation of the update equations for the Variational Bayes algorithm. The procedure is described in (Zarb Adami, 2003), (Miskin, 2000) and (Penny \& Roberts, 2000). Here it is the results from the latter approach that we use and we refer the reader to this paper for the algorithmic details. The algorithm is stopped when successive changes in the Evidence bound are less than $10^{-6}$. The plots in Figure \ref{fig:GMM3} show the results of fitting different numbers of Gaussians to the data.
The fit having 3 Gaussians results in the following values for the means and the widths of the three component distributions:
\begin{eqnarray}
\mathbf{\mu}&=&(-0.94, 1.02, 2.98) \\
\mathbf{\sigma}&=&(0.42, 0.33, 0.68)
\end{eqnarray}
These are in close agreement with the true values with which the data was generated. The deduced ratio of the three Gaussians is also correct, given by \begin{equation} 0.299:0.353:0.347 \end{equation}
The plot of the evidence bound in Figure (\ref{fig:GMM3}) peaks at the value of three, correctly indicating that most probably three Gaussians were used to construct the Mixture Model. The code takes around 25 seconds to run when testing for 1 to 6 Gaussians. 
\begin{figure}
\begin{center}
$\begin{array}{c@{\hspace{0.05in}}c}
\includegraphics[width=1.6in]{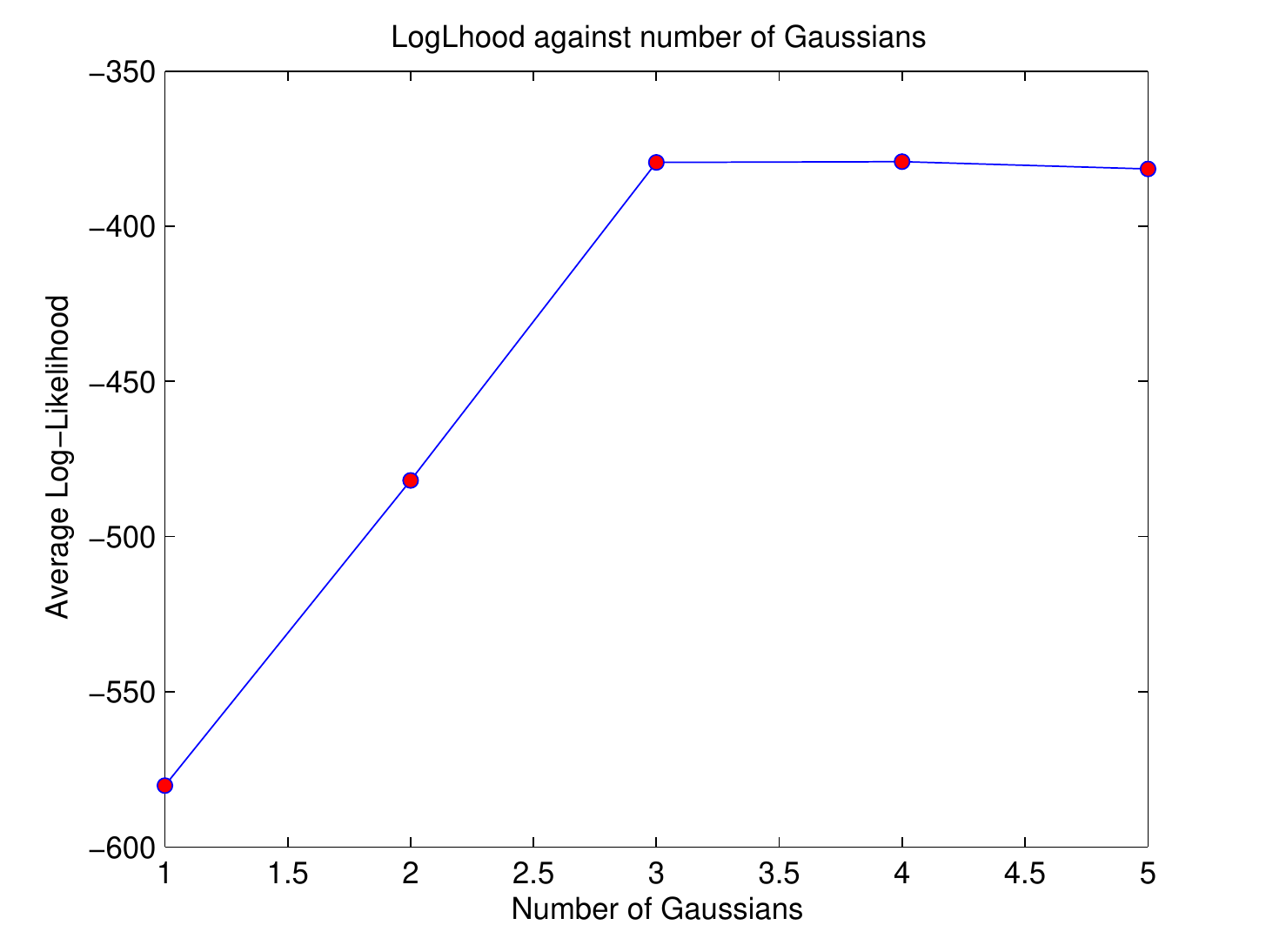}  &  
\includegraphics[width=1.6in]{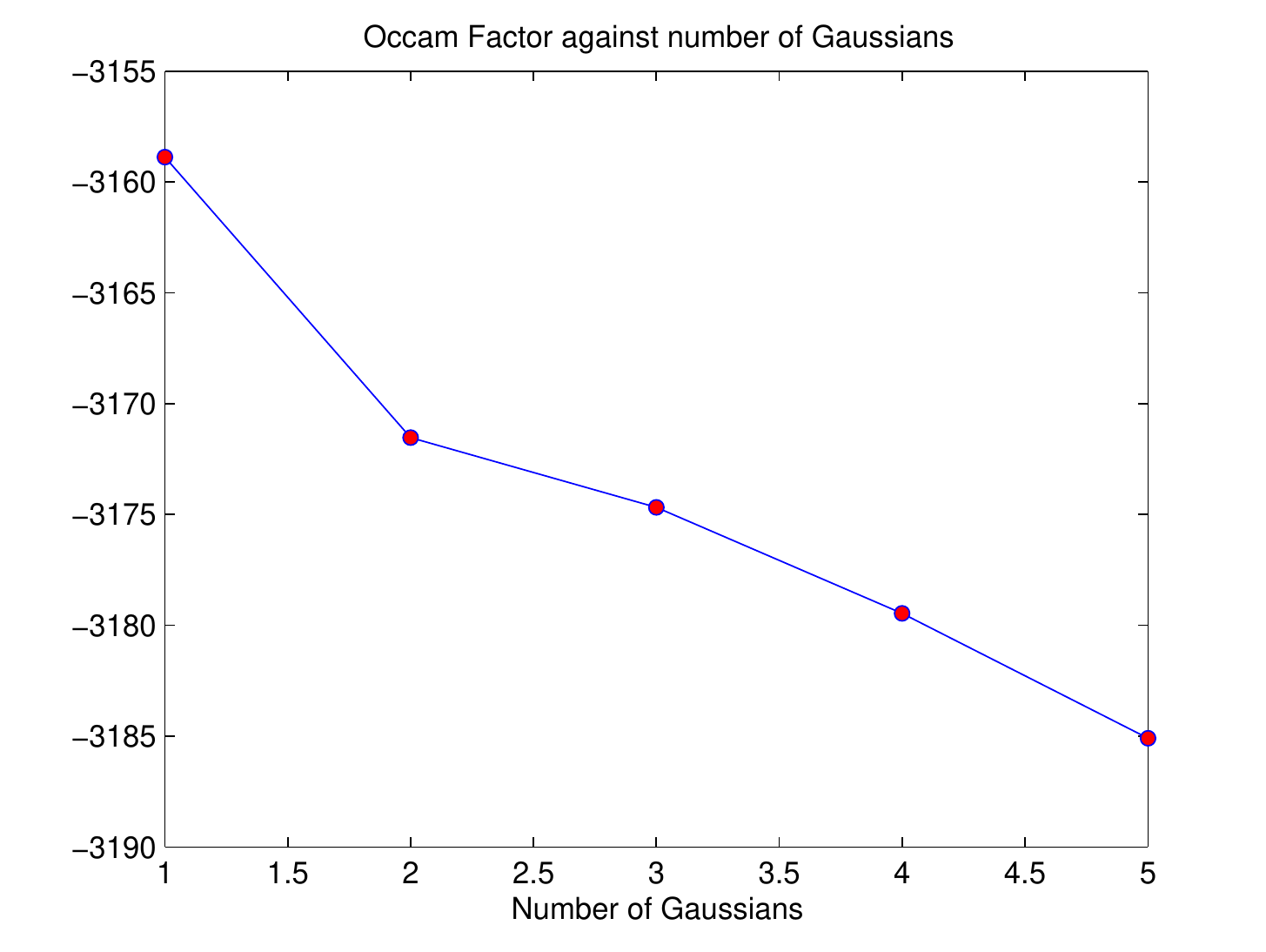}  
\end{array}$
$\begin{array}{c}
\includegraphics[width=1.6in]{./Figures/Ch4GMMEasyEvidence.pdf}  \\
\end{array}$
\end{center}
\caption[Evidence for GMM using VB]{A plot of the log-Likelihood, the Occam factor and the Evidence for different numbers of Gaussians.}
  \label{fig:GMM2}
  \end{figure}

\subsubsection{Nested Sampling}
\begin{figure}
\begin{center}
$\begin{array}{c@{\hspace{0.05in}}c}
\includegraphics[width=1.6in]{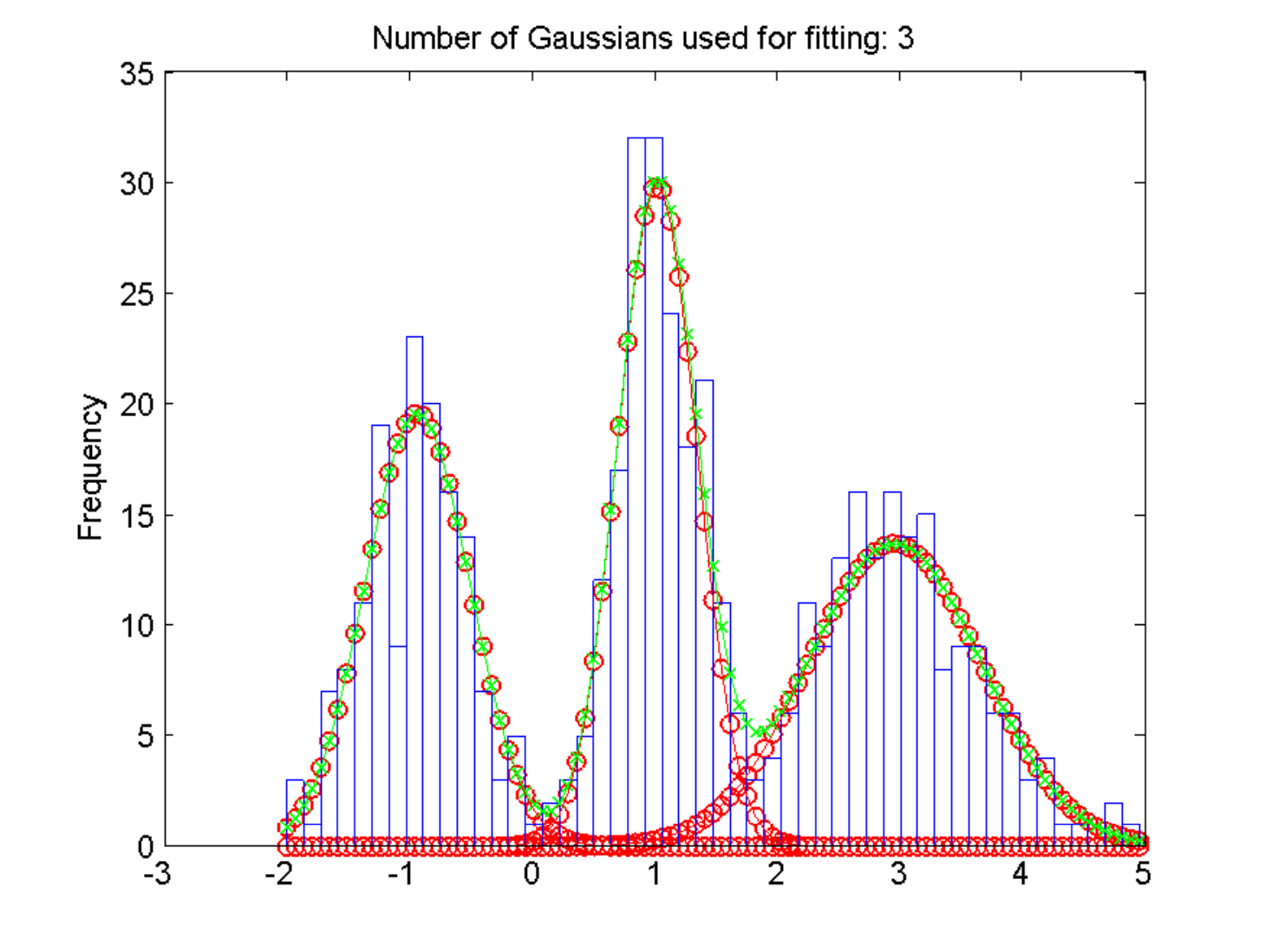}  &  
\includegraphics[width=1.6in]{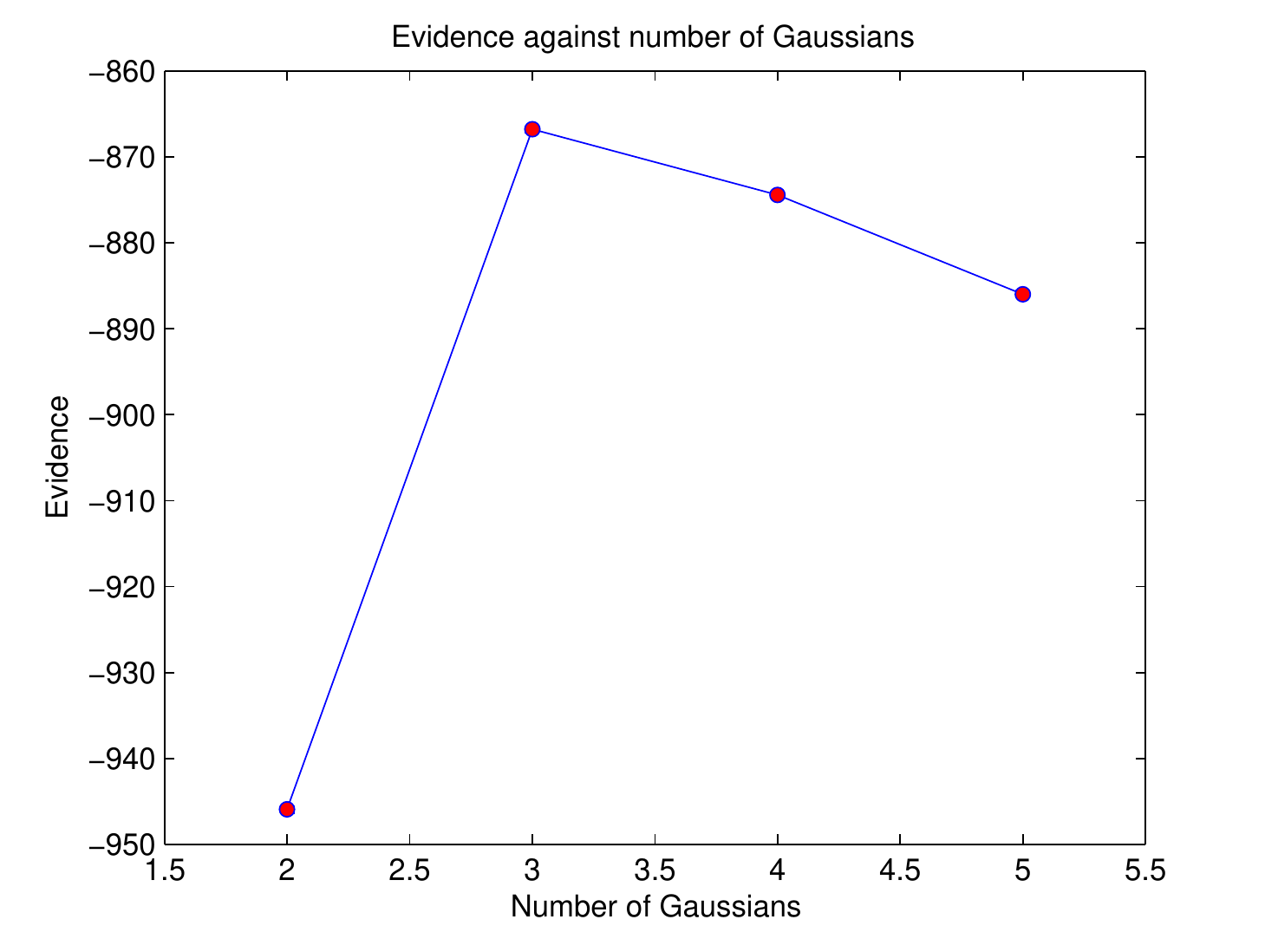} \\
\end{array}$
\end{center}
\caption[GMM Results using Nested Sampling]{GMM Evidence Plot using Nested Sampling (right) and fitted GMM (left).}
 \label{fig:GMM3NS}
\end{figure}

We now apply the Nested Sampling algorithm to tackle the Gaussian Mixture Model data. As a stopping condition we impose that the difference between successive values in $\log \mbox{Evidence}$ should be less than $10^{-6}$ and that the condition in Equation (\ref{equ:stopping}) is satisfied. In the results quoted here we utilize 50 objects in order to ensure a decent algorithmic speed. Once again the Nested Sampling implementation is considerably slower, with the code taking approximately 8 minutes to try out 1 to 6 Gaussians. Again, as with polynomial fitting, we can attribute the difference in computational speed to the random nature of the Nested Sampling algorithm. We note here that for both polynomial fitting as well as Gaussian Mixture Models the Nested Sampling implementation is our own and hence further work might be able to improve the computational speed. However, we do not believe that any increase in speed for Nested Sampling will be able to improve the computational time to Variational Bayes levels. In Figure \ref{fig:GMM3NS} we show the Evidence against the number of Gaussians used for fitting as well as the optimal fit using three Gaussians.
We first note that the figure gives a peak at the correct value of 3, indicating that 3 Gaussians were used to generate the data set. For the optimal fit the three Gaussians have means $\mathbf{\mu}= (-0.97, 1.02, 2.95)$, sigmas $\mathbf{\sigma}=(0.40, 0.35, 0.70)$ and are produced in the ratio $(0.292:0.354:0.354)$. These are very close to the true values which were used to generate the data set and in particular we note that the ratios are calculated extremely well. Differing runs of the Nested Sampling algorithm, particularly when we run with more stringent stopping conditions, result in parameter values which differ by $0.03$ at most. For a run with a stopping condition of $10^{-6}$ we get $\mathbf{\mu}= (-0.95, 1.01, 2.93)$, $\mathbf{\sigma}=(0.42, 0.32, 0.73)$ and ratio = $(0.291:0.354:0.354)$ . 

\begin{figure}
\begin{center}
$\begin{array}{c@{\hspace{.05in}}c}
\includegraphics[width=1.6in]{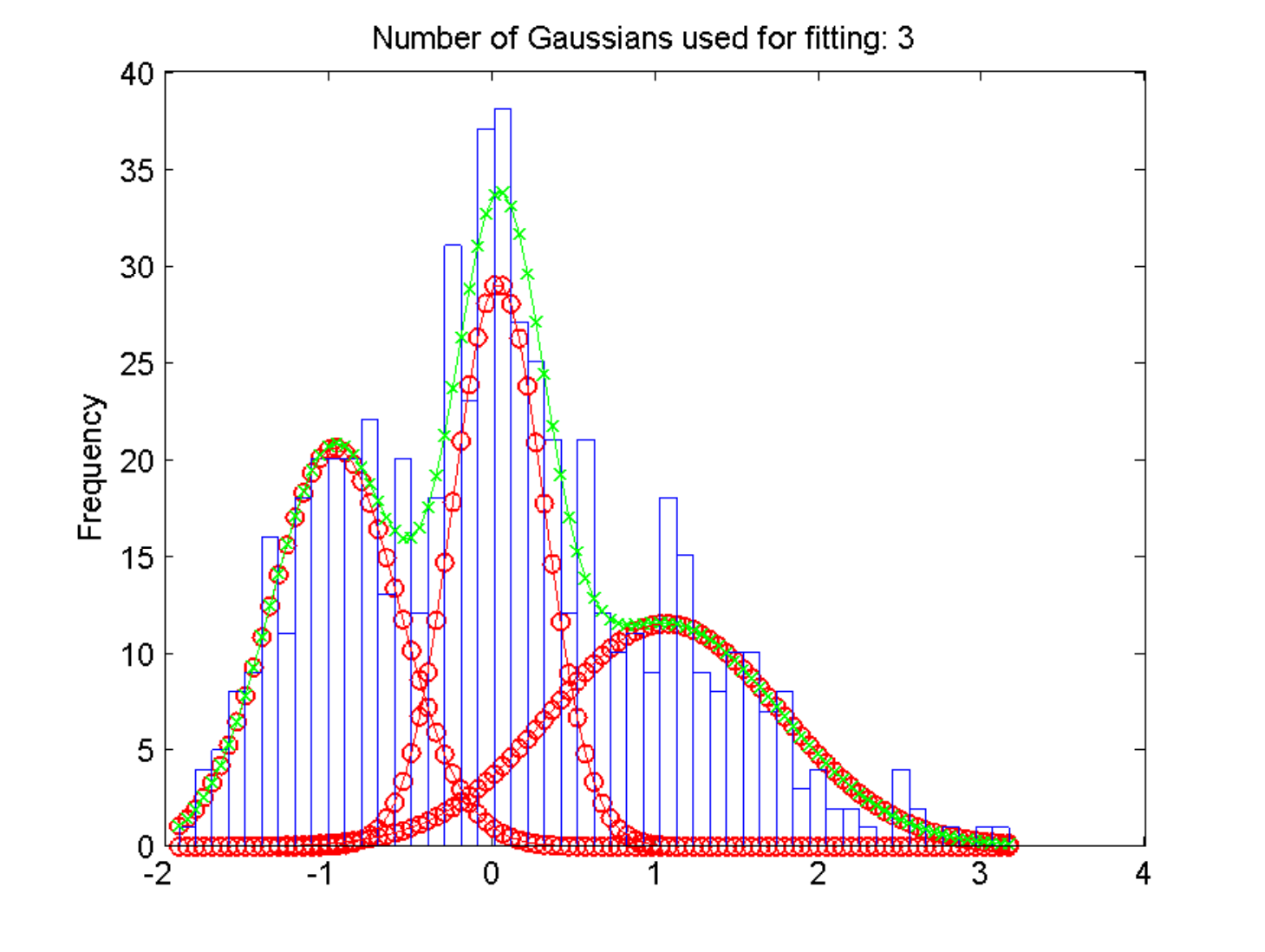}  &  
\includegraphics[width=1.6in]{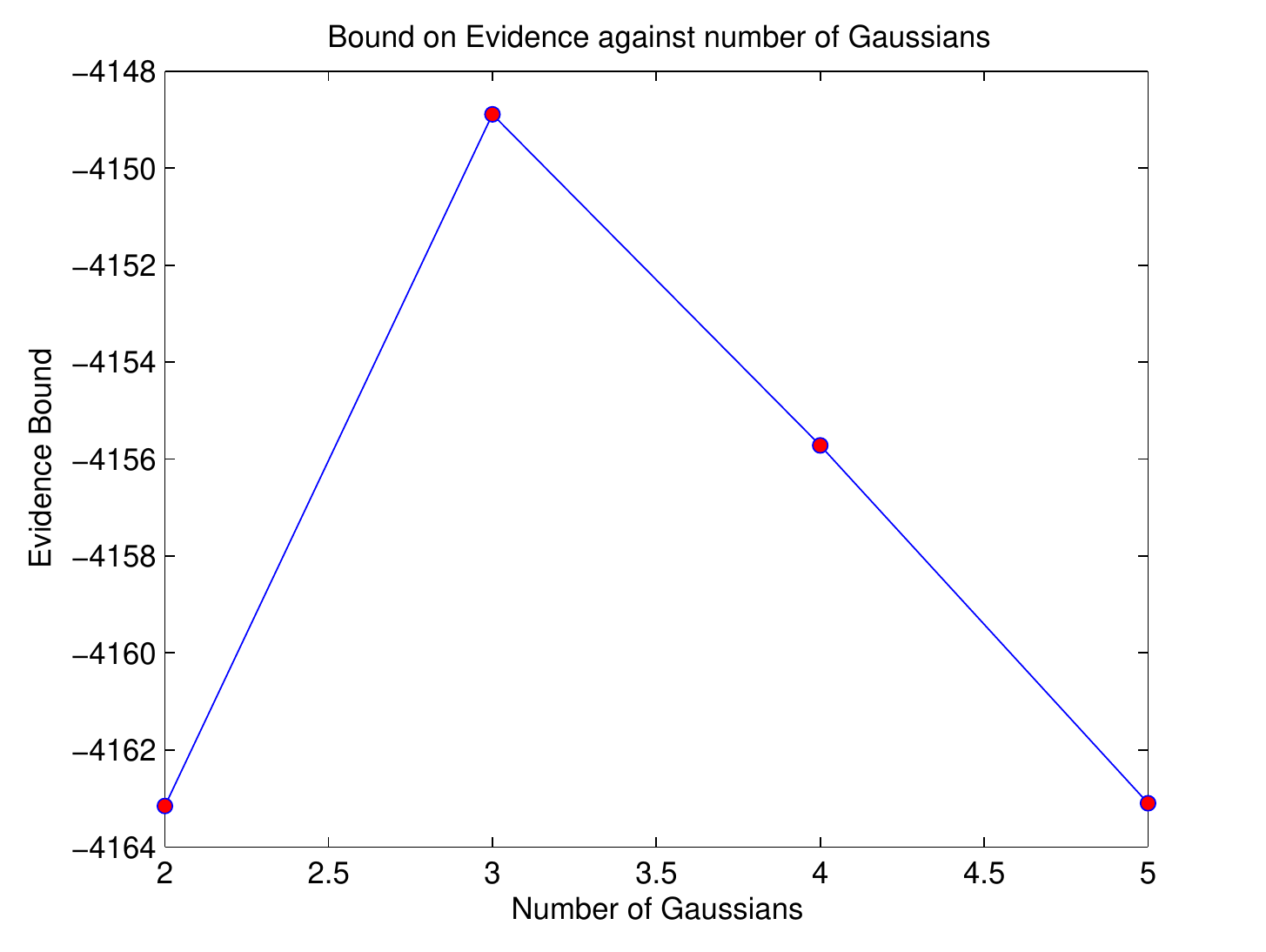} \\
\end{array}$
\end{center}
  \caption[GMM Results using Variational Bayes with a Hard data set]{GMM Results using Variational Bayes with a hard data set. On the left we see the optimal fit using three Gaussians. On the right we have a plot of the Evidence against the number of Gaussians used.}
  \label{fig:GMMHardVB}
  \end{figure}

We also test both algorithms using a harder data set where there is far more overlap in the Gaussians. To perform this analysis we generate 600 data points with means $\mathbf{\mu}=(-1,0,1)$, leaving the other parameters as before. The results are again favourable. In particular the Evidence bound peaks at the correct value of 3. Also the inferred parameters of the 3 fitted Gaussians are largely in line with the true values:
\begin{eqnarray}
\mathbf{\mu} &=& (-0.96, 0.04, 1.06) \\
\mathbf{\sigma} &=& (0.39, 0.28, 0.70)
\end{eqnarray}
The ratios are given by:
\begin{equation}
0.326 : 0.342 : 0.332
\end{equation}
The largest discrepancy is in the ratio values. This probably stems from fact that because the Gaussians are so close to each other it is harder to determine from which Gaussian each point has been generated. These results are shown in Figure \ref{fig:GMMHardVB}.
Investigating the same data set with Nested Sampling we get the following results.
\begin{figure}
\begin{center}
$\begin{array}{c@{\hspace{.05in}}c}
\includegraphics[width=1.6in]{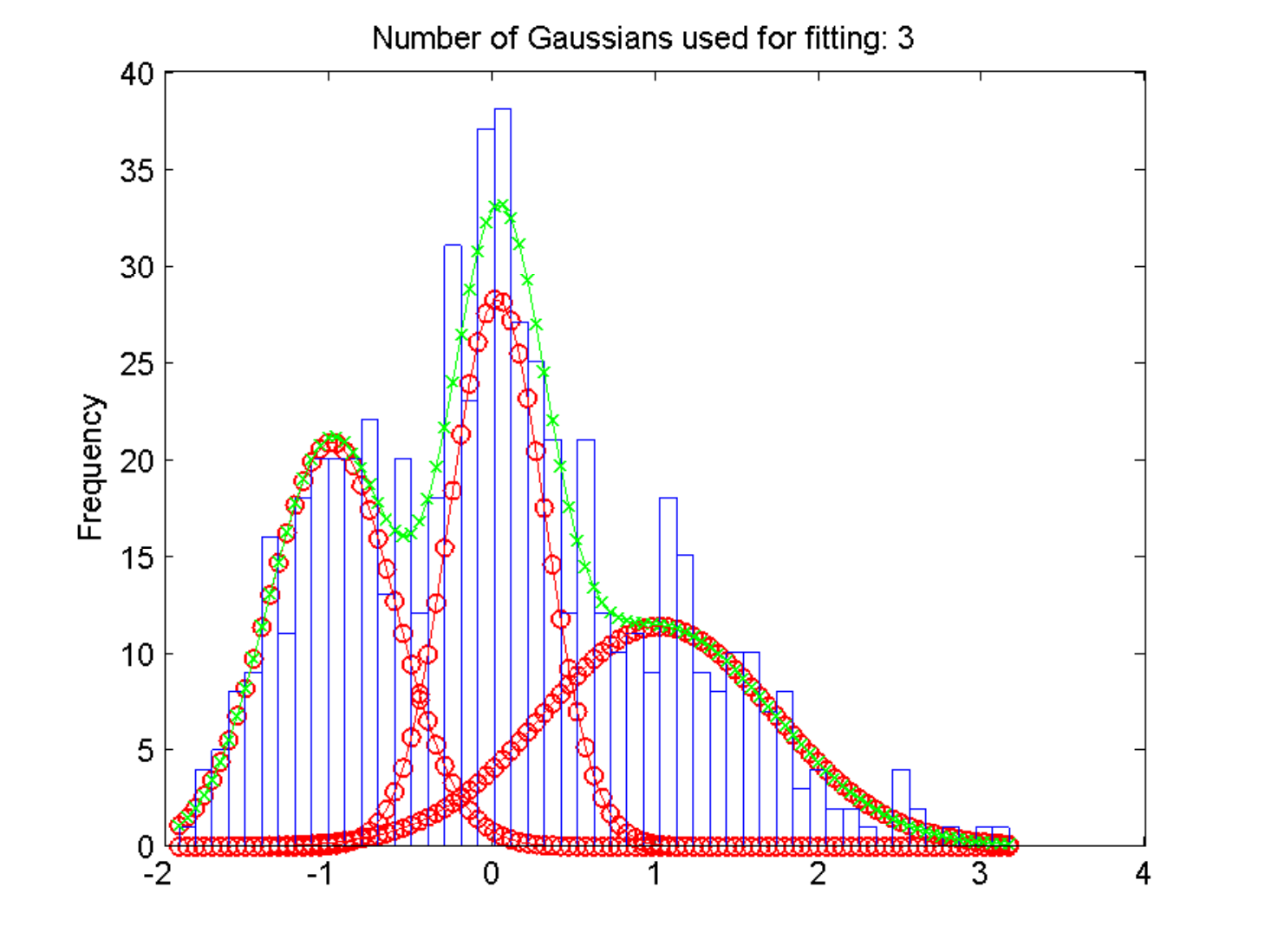}  &  
\includegraphics[width=1.6in]{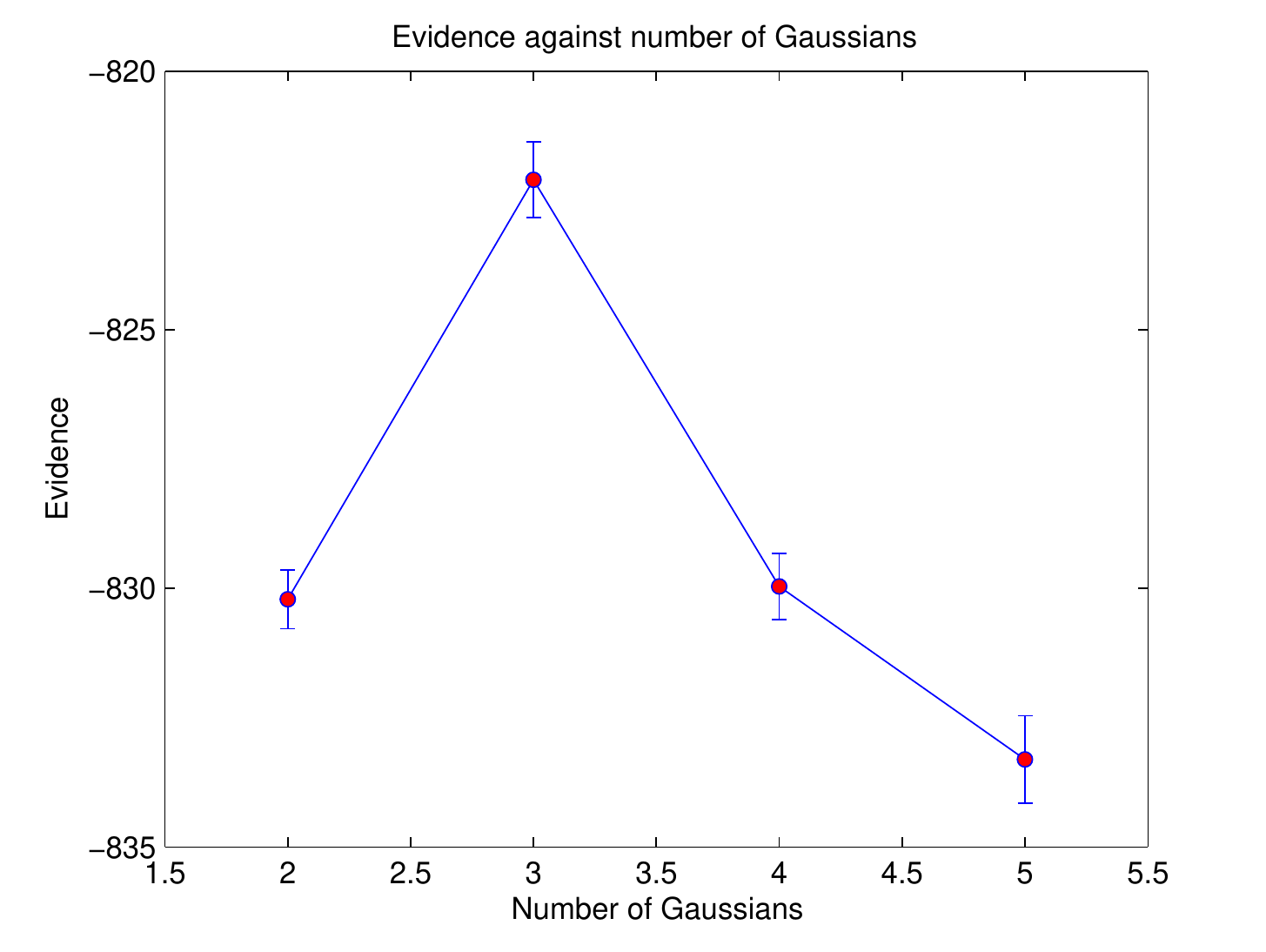} \\
\end{array}$
\end{center}
  \caption{GMM Results using Variational Bayes with a hard data set. On the left we see the optimal fit using three Gaussians. On the right we have a plot of the Evidence against the number of Gaussians used.}
  \label{fig:GMMHardNested}
  \end{figure}
The Gaussian positions returned by the algorithm are given by:
\begin{eqnarray}
\mathbf{\mu} &=& (-0.98,    0.03 ,     1.02)\\
\mathbf{\sigma} &=& (0.39,     0.29,     0.70)   \\
\mbox{ratios} &=&  (0.331 :     0.340 :    0.329)  
\end{eqnarray}
Nested sampling produces slightly better values for the Gaussian means and widths but the ratios are again incorrect. The values are however within one standard deviation of the true results. Again there is good agreement between the Variational Bayes and Nested Sampling methods and differences in the results from the true values are reflected in both techniques. This indicates that much of the imperfections in the inference are due to the random nature of the generated data and not due to the algorithmic deficiencies.

\section{Conclusions}

Virtually all experimental physical analysis done today is somewhat statistical in nature and therefore necessitates the use of powerful computational tools. The Bayesian framework is one such set of tools which is gaining popularity within the scientific community. Our main conclusion is that whilst both Variational Bayes and Nested Sampling give similar and accurate results when tackling engineered problems the former is much faster than the latter. Inferred parameters using the two different techniques are often within a few percent of each other and the peaks in the Evidence plots always agree. The discrepancy in speeds was attributed to the fact that Nested Sampling is an MCMC method and is dependent on random exploration of a parameter space while Variational Bayes is analytical and has no such probabilistic element. We do however point out that it is far easy to formulate a Bayesian solution to a problem using Nested Sampling than using Variational Bayes as in the latter case one must derive the update equations for each problem and this might be a non-trivial process.

\subsection*{Acknowledgments}
I would like to thank the Square Kilometre Array Design Studies (SKADS) for funding my stay at the University of Oxford in July 2009 where I finished off this work.\newline

\appendix

\subsection*{Appendix A: Notation}

We describe the distributions used throughout this paper as well as their notation.
\newline
\textbf{Gaussian Distribution}
\newline
The Gaussian (or Normal) distribution is given by the Equation \ref{equ:gauss}, where we consider all previous knowledge to be encoded in $\mathbb{I}$
\begin{eqnarray}
P(x|\mathbb{I})&=&G(x|\hat{x}, \tilde{x}) \\
\label{equ:gauss}
&=& \sqrt{\frac{\tilde{x}}{2\pi}}\exp\left(-\frac{\tilde{x}}{2}(x-\hat{x})^2)\right)
\end{eqnarray}
Here $\hat{x}$ is the mean and $\tilde{x}>0$ is the inverse variance. Hence the standard deviation, $\sigma$ given by the square root of the variance can be expressed as follows:
\begin{equation}
\sigma = \sqrt{\mbox{variance}}=\sqrt{\frac{1}{\tilde{x}}}
\end{equation}
Useful expectation values under the Gaussian distribution are: 
\begin{equation}
<x>_P=\hat{x} \quad <x^2>_P=\hat{x}^2+\tilde{x}^{-1}
\end{equation}

\textbf{Multivariate Gaussian Distribution}
\newline
The Gaussian can be generalised to $d$ dimensions, in which case it is called a Multivariate Gaussian:
\begin{eqnarray}
P(\bm{x}|\mathbb{I})&=&G(\bm{x}|\bm{\hat{x}}, \bm{\tilde{x}}) \\
&=& \sqrt{\frac{\bm{\tilde{x}}}{(2\pi)^d}}\exp\left(\frac{-1}{2}(\bm{x}-\bm{\hat{x}})^T\mathbf{\tilde{x}}(\mathbf{x}-\mathbf{\hat{x}})\right)
\end{eqnarray}
Here $\mathbf{\hat{x}}$ is a vector containing the mean in each dimension and $\mathbf{\tilde{x}}>0$ is the symmetric and positive definite inverse covariance matrix. This is necessarily a symmetric matrix because the cross terms are related to the covariances of the variables and the covariance relation is necessarily a symmetric one. When the variables are independent the covariance matrix, and hence the inverse covariance matrix, is of diagonal form. Useful expectation values under the multivariate Gaussian distribution are:
\begin{equation}
<\bm{x}>_P=\hat{\bm{x}} \quad <\bm{x^2}>_P=\hat{\bm{x}}^2+\tilde{\bm{x}}^{-1}
\end{equation}
\textbf{Gamma Distribution}
The Gamma distribution crops up in the Bayesian framework because it is the conjugate distribution for the inverse variance of a Gaussian distribution. The distribution itself is given by:
\begin{eqnarray}
P(\bm{x}|\mathbb{I})&=&\mbox{Gamma}(\bm{x}|a,b) \\
&=& \frac{1}{\Gamma(b)}a^bx^{(b-1)}\exp(-ax)
\end{eqnarray}
The constant $a>0$ is a scale variable and $b>0$ dictates the shape of the distribution. Some useful expectation values are:
\begin{eqnarray}
<x>_P&=&\frac{b}{a}\\
<\log x>_P &= &-\log a +\frac{\partial \log\Gamma(b)}{\partial b} 
\end{eqnarray}
The derivative $\frac{\partial \log\Gamma(b)}{\partial b} $ is known as the \emph{digamma} function and is often denoted by $\Psi(b)$.
\newline\textbf{Dirichlet Distribution}
The Dirichlet distribution is used to model categorical weights, particularly in the Gaussian Mixture Model setting. The probability distribution function for a Dirichlet describing $m$ weights is given by:
\begin{equation}
\displaystyle
P(\pi)=\frac{\Gamma(\sum_{s=1}^m\lambda_s)}{\prod^m_{s=1}\Gamma(\lambda_s)}\prod_{s=1}^{m}\pi_s^{\lambda_0-1}
\end{equation}
The parameter $\lambda_0$ is called the mixing hyperparameter. The word hyperparameter is often used in Bayesian statistics to denote parameters in priors and to distinguish them from the actual parameters of the underlying model under investigation. When the $\lambda_s$ are all equal we get the symmetric Dirichlet distribution:
\begin{equation}
\displaystyle
P(\pi)=\frac{\Gamma(m\lambda_0)}{\Gamma^m(\lambda_s)}\prod_{s=1}^{m}\pi_s^{\lambda_s-1}
\end{equation}
The expectation value of $\pi_i$ is given by:
\begin{equation}
\displaystyle
<\pi_i>_P=\frac{\lambda_i}{\sum_{i=1}^{m}\lambda_i}
\end{equation}

\bsp

\label{lastpage}

\end{document}